\documentclass[fleqn,usenatbib]{mnras}

\usepackage{newtxtext,newtxmath}

\usepackage[T1]{fontenc}

\DeclareRobustCommand{\VAN}[3]{#2}
\let\VANthebibliography\thebibliography
\def\thebibliography{\DeclareRobustCommand{\VAN}[3]{##3}\VANthebibliography}

\usepackage{graphicx}	
\usepackage{amsmath}	

\usepackage{anyfontsize}

\title[The Unified Cluster Catalogue]{The Unified Cluster Catalogue: towards a
comprehensive and homogeneous database of stellar clusters}

\author[G. I. Perren et al.]{
Gabriel I. Perren,$^{1,3}$\thanks{E-mail: gabrielperren@gmail.com}
Mar\'ia S. Pera,$^{2,3}$
Hugo D. Navone$^{2,3}$
and Rub\'en A. V\'azquez$^{1,4}$
\\
$^{1}$Instituto de Astrof\'isica de La Plata, IALP (CONICET-UNLP), 1900 La Plata, Argentina\\
$^{2}$Instituto de F\'isica de Rosario, IFIR (CONICET-UNR), 2000 Rosario, Argentina\\
$^{3}$Facultad de Ciencias Exactas, Ingenier\'ia y Agrimensura (UNR), 2000 Rosario, Argentina\\
$^{4}$Facultad de Ciencias Astron\'omicas y Geof\'isicas (UNLP), 1900 La Plata, Argentina
}

\date{Accepted XXX. Received YYY; in original form ZZZ}

\pubyear{2023}

\begin{document}
\label{firstpage}
\pagerange{\pageref{firstpage}--\pageref{lastpage}}
\maketitle

\begin{abstract}
We introduce the Unified Cluster Catalogue, the largest catalogue of
stellar clusters  currently listing nearly 14000 objects.
In this initial  release it exclusively contains Milky Way open
clusters, with plans to include other objects in future updates.
Each cluster is processed  using a novel probability membership
algorithm,  which incorporates the coordinates, parallax,
proper motions, and their  associated uncertainties  for each
star into the probability assignment process.
We employ Gaia DR3 data up to a G magnitude of 20, resulting in  the
identification of over a million probable members. The catalogue is
accompanied by a  publicly accessible website designed to simplify
the search and data exploration of stellar clusters.  The website can be
accessed at \url{https://ucc.ar}.
\end{abstract}

\begin{keywords}
(Galaxy:) open clusters and associations: general --
catalogues -- methods: data analysis
\end{keywords}



\section{Introduction}

Open clusters (OCs) are groups of  loosely gravitationally bound coeval
stars with a wide range of masses, formed from the same molecular cloud.
Their orbits generally  position them close to the  formal
Galactic plane, although there are a few examples of OCs with large vertical
distances. Having originated from the same cloud, their member stars share a
 common chemical composition and age, and are located  within a
somewhat compact  region in space.
The study of OCs  holds fundamental importance for  several key
aspects in astrophysics research, including the process of stellar evolution as
well as the dynamics, structure, formation, and chemical evolution of the
 Galactic disk~\citep{Friel1995}.

Catalogues of OCs are  essential for organizing these objects into publicly
available databases and, most importantly, as a tool for helping researchers
discover new potential OCs while discarding false detections.
Catalogued bonafide OCs are the most natural training dataset against which
we can test and improve new algorithms for stellar clusters detection.
There have been efforts to provide catalogues of OCs to the astrophysical
research community at least since the late 18th century, even if that was not
the primary objective of the compilation. The first of such well known
catalogues to include OCs is the Messier Catalogue~\citep{Messier1771} with
less than 30 OCs listed. This work was quickly followed by Herschel's
Catalogue of One Thousand New Nebulae and Clusters of
Stars~\citep{Herschel1786}, culminating a century later with Dreyer's New
General Catalogue of Nebulae and Clusters of Stars~\citep[NGC,][]{Dreyer1888}.
The NGC lists less than 650 OCs, which is more than twenty times the number of
objects present in Messier's catalogue.

After that first rapid growth the pace with which OCs were discovered and
catalogued slowed down. The next big catalogue, again published almost a
century later, was the Base Données Amas~\citep{Mermilliod1995} which
listed a little over 1100 OCs, based on the previous compilation
by~\cite{Lynga1987}. This work is the foundation of the WEBDA
catalogue,\footnote{\url{https://webda.physics.muni.cz/webda.html}} a heavily
used resource in the analysis of OCs.
In the following decade, with the advent of large public databases containing
from hundreds of thousands to millions of stars ( e.g., Hipparcos and
Tycho~\citep{Perryman_1997,Hog_1997}, as well as 2MASS~\citep{Skrutskie_2006}),
the task of detecting new candidate OCs became substantially more attainable.
This is particularly true for those objects that are faint, obscured by dust, or
not in the vicinity of the solar system. Catalogues such as the Milky Way Star
Clusters~\citep{Kharchenko_2012} or those presented
by~\cite{Loktin_2017} or~\cite{Bica_2019} increased the number of  known
OCs to more than 3000 in a few decades.

\begin{figure}
	\includegraphics[width=\columnwidth]{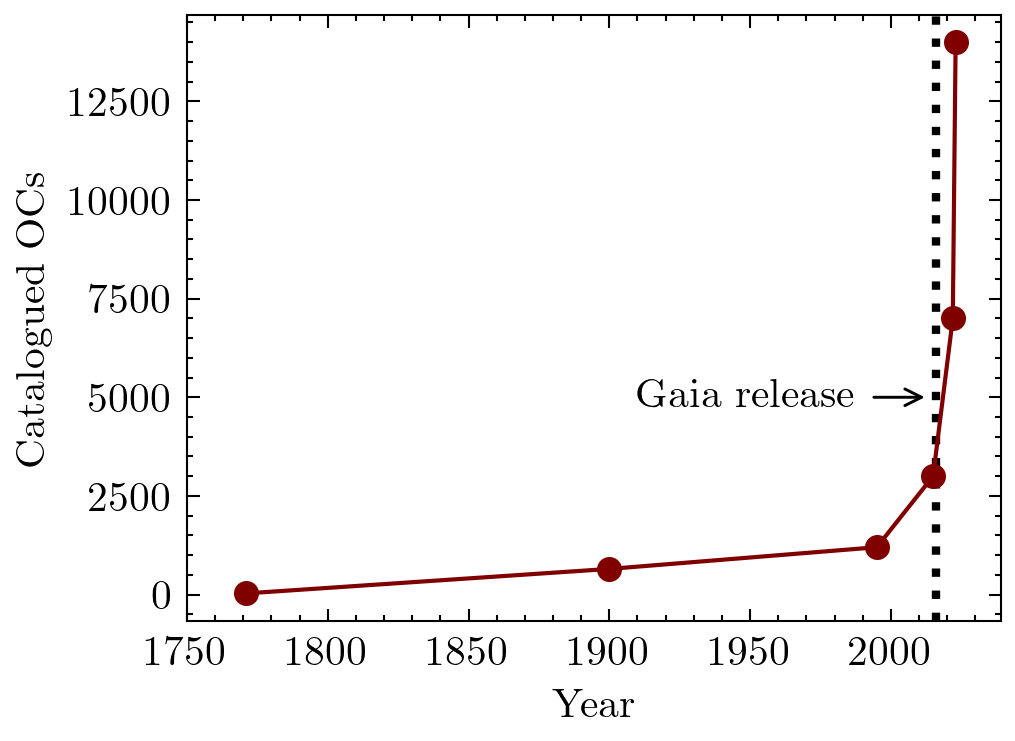}
    \caption{Approximate number of catalogued OCs in the literature 
     from the late 1700s to the present day, including this work.
    The release date of Gaia's survey data is  indicated with a dotted
    line.}
    \label{fig:catalogued_ocs}
\end{figure}

 Currently, the release of the Gaia survey database~\citep{Gaia_2016} with over a
billion observed stars  has led to an enormous quantity of new candidate
OCs being reported in the literature.
Clustering algorithms such as HDBSCAN~\citep{Campello_2013} are used for the
automatic identification of overdensities,  dramatically improving the
detection sensitivity compared to manual methods.
The latest catalogue  that made use of both Gaia data and this
algorithm is the one  published in~\cite{Hunt_2023}.
 It contains a little over 6000 OCs, $\sim$2000 of which are new
candidates.
 Just the current year, at the moment of writing this article,
approximately 5000 new candidate OCs have been presented.
 In Fig.~\ref{fig:catalogued_ocs} we show the growth in catalogued
OCs  over the last two and a half centuries.  The last two
decades show an almost exponential trend,  with no signs of slowing
down. Estimates for the total number of OCs in the Galaxy locate the upper limit
over 10$^{5}$~\citep{Bonatto_2006}, which means that there remain still a lot of
objects waiting to be found.

 Studies where thousands of new candidate OCs are introduced present
a real challenge. Depending on the  input parameters  of the
clustering algorithms employed  in these works, apparent groupings of
non-physically related stars can easily disguise themselves as OCs.
 Such false detections are prone to remain hidden in these catalogues
since the number of objects is too large to  allow a
detailed individual analysis.
%
%
%
Furthermore,  authors are often unable to cross-reference their
findings with other recent catalogues, due to the dispersion of information
across articles and the influx of new articles published just months apart.
This translates into duplication issues which, combined with the aforementioned
problem, can have non trivial effects when these objects are used in massive
studies (for example, when analysing the Galactic  disk structure).

Our aim is to provide the community with a service that can be used to
alleviate these issues. We combined, to the best of our knowledge, all the
recent catalogues of OCs in the literature, and used them to generate a single
catalogue of $\sim$14000  unique OCs. We call this the Unified Cluster
Catalogue (UCC hereafter).
 The UCC contains several thousands more objects than similar services
such as WEBDA\footnote{\url{https://webda.physics.muni.cz/}} and
SIMBAD,\footnote{\url{https://simbad.u-strasbg.fr/simbad/}} which list
$\sim$1800 and $\sim$5200 objects, respectively.
Moreover, while neither WEBDA nor SIMBAD provide homogeneous membership
information, we applied a new tool called \texttt{fastMP}\footnote{\texttt{fastMP}:
\url{https://github.com/Gabriel-p/fastMP}} to assign membership probabilities to
each listed OC.
The code was designed to work on the Gaia survey data, resulting
so far in more than 1 million stars identified as probable members of the
catalogued OCs. This gives us the largest homogeneously processed
database of OC members to date. The data collected for each OC in the UCC
catalogue, along with their estimated members, is publicly available in the
website \url{https://ucc.ar}. This website will  undergo periodic
updates as new candidate OCs are introduced to the community.

This article is structured as follows. In Sect.~\ref{sec:databases} we
introduce the stellar cluster databases employed to generate this initial
version of the  UCC. Section~\ref{sec:methods} presents the
new membership algorithm, \texttt{fastMP}, employed in the study of all the
catalogued clusters.
A comparison of our membership results with those recently published is
performed in Sect.~\ref{sec:results}, along with the presentation of the public
web site where these results will be hosted, updated, and expanded in the future.
Finally, our conclusions are highlighted in Sect.~\ref{sec:conclusions}.

\section{Databases}
\label{sec:databases}

We gathered the OCs listed in the literature up to  the present day, extracted
from 32 databases published in the last 11 years. The total number of entries is
24983,  which were cross-matched to  create a final catalogue of
13684 unique OCs.
A few small databases were not added as they are included entirely
in~\cite{Hunt_2023},  these are: \cite{Zari_2018}, \cite{Bastian_2019}, 
\cite{Tian_2020}, \cite{Qin_2021}, \cite{Anders_2022}, \cite{Casado_2023}.
Table~\ref{tab:references} lists these  32 databases along with the
number of OCs that were left  in each, after extensive cleaning and
sanitizing. To avoid cluttering the article we  defer the details 
 of these cleaning procedures to Appendix~\ref{app:db_cleaning}, and
restrict this section for a general discussion of the  cross-matching
process.\\

\begin{table}
	\centering
	\caption{ List of all the catalogues  cross-matched in this
	article to generate the	 UCC database. Columns ID and N are the
	denomination used for new candidate OCs, and the
	total number of OCs	taken from  each work, respectively.}
	\label{tab:references}
	\begin{tabular}{lcr}
		\hline
		Reference & ID & N\\
		\hline
		\cite{Kharchenko_2012}  & -- & 2854\\
		\cite{Loktin_2017}   & LP & 1050\\
		\cite{Castro_Ginard_2018} & UBC   & 23\\
		\cite{Bica_2019} &  Bica  & 3555\\
		\cite{Castro-Ginard_2019}  & UBC & 53\\
		\cite{Sim_2019}  & UPK    & 207\\
		\cite{Liu_2019}  & FoF    & 76\\
		\cite{Ferreira_2019}  & UFMG   & 3\\
		\cite{Castro-Ginard_2020}  & UBC  & 570\\
		\cite{Ferreira_2020}  & UFMG  & 25\\
		\cite{Cantat-Gaudin_2020} & --  & 2017\\
		\cite{Hao_2020}   & HXWHB  & 16\\
		\cite{Ferreira_2021}  & UFMG   & 34\\
		\cite{He_2021}  & HXHWL  & 74\\
		\cite{Dias_2021}  &  --  & 1742\\
		\cite{Hunt_2021}  & PHOC  & 41\\
		\cite{Casado_2021}  & Casado  & 20\\
		\cite{Jaehnig_2021}   & XDOCC   & 11\\
		\cite{Santos-Silva_2021}  & CMa   & 5\\
		\cite{Tarricq_2022}   & --  & 467\\
		\cite{Castro-Ginard_2022} & UBC  & 628\\
		\cite{He_2022}  & CWNU  & 541\\
		\cite{He_2022_1}   & CWNU  & 836\\
		\cite{Hao_2022}  & OC   & 703\\
		\cite{Li_2022}  & LISC   & 61\\
		\cite{He_2023}  & CWNU   & 1656\\
		\cite{Hunt_2023}   & HSC  & 6272\\
		\cite{Qin_2023}  & OCSN    & 101\\
		\cite{Li_2023}  & LISC  & 35\\
		\cite{Chi_2023_2} & CWWL  & 46\\
		\cite{Chi_2023}  & LISC-III    & 82\\
		\cite{Chi_2023_3}  & CWWDL  & 1179\\
		\hline
		Number of OCs in all the catalogues    &   & 24983\\
		Number of unique OCs after cross-matching  &  & 13684\\
		\hline
	\end{tabular}
\end{table}

Unlike other works, such as the recent \citet[][HUNT23]{Hunt_2023}, we do not
attempt to cross-match OCs  present in different databases using 
 their positions
or astrometry. Instead, we standardize the names of  the OCs
 in the 32 selected databases, and  use these names to find
matching entries across all of them.
Although we do not attempt to find duplicate OCs via positions or astrometry
 cross-matching, we do use this data to  identify and flag the
most obvious duplicate entries. For example, the CWWDL 14677
candidate OC presented in~\cite{Chi_2023_3} has coordinates and astrometry
values estimated as: RA=0.9526\textdegree, DEC=-30.002\textdegree, pmRA=4.222 
mas yr$^{-1}$, pmDE=18.721 mas yr$^{-1}$, plx=2.61 mas. This is marked in the
UCC as a duplicate of the well known OC Blanco 1, which has almost identical
coordinates and astrometry values in the literature: RA=0.9149\textdegree,
DEC=-29.958\textdegree, pmRA=4.215 mas yr$^{-1}$, pmDE=18.724 mas yr$^{-1}$, and 
plx=2.59 mas. There are many entries with similar issues  throughout
the databases used to generate the UCC.
Note that although cases such as this one are flagged as possible duplicates,
they are not removed from  our final catalogue.
The  rationale behind this  decision is that the authors
 originally introduced these objects as new  discoveries, and we
 choose to retain this nomenclature.
Even  when duplications are evident, we believe this decision can help
future research in identifying structures that were  incorrectly
labelled as new OCs,  thereby preventing the repetition of the same error.
 Labelling duplicates can also aid in the detection and analysis of
binary cluster systems.
We will show in Sect.~\ref{sec:results} how each OC is assigned a probability of
being a duplicate of others in the  UCC catalogue, based on a simple set of
rules.  These rules, even if  somewhat arbitrary, represent a
reasonable initial step toward a more in-depth analysis of duplicate candidate
OCs in the literature.

\begin{figure*}
	\includegraphics[width=\textwidth]{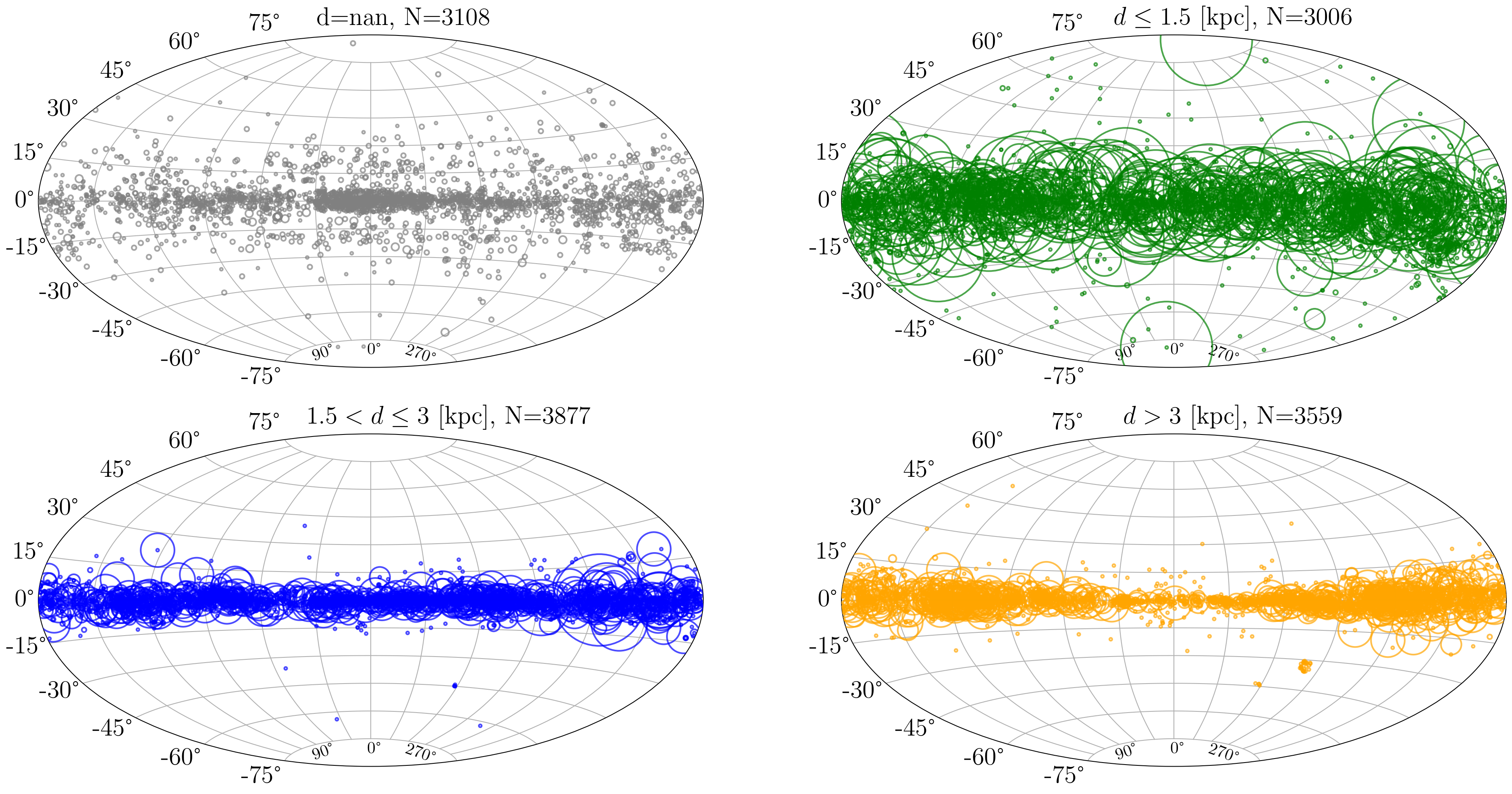}
    \caption{Map of the full  database of catalogued OCs in this work.
     Positions displayed in galactic coordinates,  segregated by
    range of catalogued distance. From top left
    to bottom right: OCs catalogued with no distance (grey), OCs with
    $d\leq1.5$ kpc (green), OCs in the range $1.5<d\leq3$ kpc (blue), and OCs
    with $d>3$ kpc (orange).  The size of each candidate OC is
    proportional to the number of databases from Table~\ref{tab:references}
     where it is included.}
    \label{fig:galactic_map}
\end{figure*}

In Fig.~\ref{fig:galactic_map} we show the distribution of the 13684 candidate
OCs catalogued in the UCC  to date, segregated by distance range. These
distances are estimated as the inverse of the  provided parallaxes. It
is worth
noting that many of the OCs in older catalogues, such
as~\cite{Kharchenko_2012} and~\cite{Bica_2019}, have
no associated astrometry but can have distances estimated as part of the
fundamental parameters analysis process (usually along with age and extinction).
As  illustrated in the upper-left plot, these catalogued OCs with no
astrometry represent almost a quarter of the  entire database.
Expectedly, OCs catalogued  at distances closer than $\sim$1.5 kpc 
(as shown in the upper-right plot) are  included in a
significantly larger number of databases,  indicated by the
proportional sizes of the circles.

The UCC includes candidate OCs found through  the analysis of infrared
photometry, such as the FSR~\citep{Froebrich_2007}, Ryu~\citep{Ryu_2018}, and
VVV~\citep{Barba_2015} objects. These  candidates are not included in
any of the recent large scale  catalogues
like~\citet[][CANTAT20]{Cantat-Gaudin_2020} and HUNT23
because they are considered to be too  faint to be detected by the usual
clustering algorithms, using Gaia photometry. A large portion of the  candidate
OCs with no  assigned distance seen in the upper-left panel of
Fig.~\ref{fig:galactic_map} are precisely these objects.
Our method, as we will  demonstrate in Sect.~\ref{sec:methods}, does not
 rely on the  capability of a clustering algorithm to detect
faint and small overdensities, which it will most likely not be able to
accomplish. Thus, we can include these  candidate OCs  in the
UCC, and  provide a first estimation of their mean positions in 
 terms of proper motions and parallax.

In this initial version of the UCC we did not include information on candidate
OCs identified  as probable asterisms,  presented, for instance,
in~\cite{Cantat-Anders_2020}.
 The rationale behind this omission is the considerably dispersed nature
of such information, rendering its collection more challenging compared to, for
example, positional data or fundamental parameter values.
We  intend to incorporate this information in future updates to the
catalogue.
 We do however assign two different quality  classification
parameters to each object, which  can be used to identify likely
non-clusters. This will be discussed in more detail in Sect~\ref{sec:results}.

Recently,~\cite{Kounkel_2020} presented a list of more than 8000 moving
groups. Many of these are very extended and small groups which
can hardly be classified as OCs. For this initial version of the UCC, we
 decided to remove these groups from the catalogues where they are
included, i.e. \cite{He_2022_1} and HUNT23.\\
While we were  in the process of preparing this manuscript, a new work
by~\cite{He_2023_2}  introduced $\sim$2000 new candidate OCs,  for which
their associated data is  not yet accessible.
Although  these candidate OCs are not included in 
 the current version of the UCC, it  is highly probable that they
will have been incorporated by the time this article is published.\\

We  utilized the most recent release of the Gaia survey
data~\citep[DR3;][]{GaiaDR3_2022,Babusiaux_2022}  implementing a single 
 filter with a maximum magnitude  cutoff at G=20 mag. No 
 additional filters were applied on this dataset, which was
employed  for the processing of the entire catalogue, extracting the
most probable members for each listed OC. In Sect.~\ref{sec:methods}
 we elaborate on this process of membership probability estimation.


\section{Membership method}
\label{sec:methods}

Once the  databases listed in Table~\ref{tab:references} are
cross-matched and the  final catalogue  with unique entries is
generated, our next objective is to compile a homogeneous database of likely
member stars for each  unique candidate OC.
 Typically, tools  known as clustering algorithms are employed for
this task. Three of the most  frequently used  clustering algorithms
in the stellar cluster literature are
Friends-of-Friends~\citep[][FoF]{Huchra_1982},
DBSCAN~\citep{Ester_1996} and HDBSCAN.
Examples of recent articles  listed in Table~\ref{tab:references}
 where these algorithms are  employed are \cite{Liu_2019}, 
\cite{He_2023}, and HUNT23, for FoF, DBSCAN, and HDBSCAN, respectively.
More specialized tools such as UPMASK~\citep{Krone_2014} or our own
recently developed pyUPMASK~\citep[][a generalized Python-based
version of UPMASK]{Pera_2021}, also depend at their cores on these clustering
algorithms.\footnote{UPMASK depends on the
K-Means~\citep{MacQueen_1967,Lloyd_1982} algorithm, while pyUPMASK is able to
work with about a dozen different clustering algorithms.}
 While these tools have been used  extensively in the
literature, we believe they  have some important shortcomings that need
to be addressed.
The first one is the processing time.
For  the amount of data we are handling,  efficient code is crucial.
For example, it took HUNT23 eight days of runtime on 48 CPU cores to process the
Gaia DR3 database using HDBSCAN. This large requirements can easily become an
obstacle in the analysis.
Second, and tightly related to the first problem mentioned previously, these
algorithms do not take uncertainties into account.
One could incorporate uncertainties associated with the input data through  a
bootstrapping mechanism~\citep{Efron_1979}, but  the initial problem would still
persist.
If a single  data processing run is  very time-consuming,
 as shown above with the HDBSCAN example in HUNT23,  conducting
thousands of runs  becomes virtually impossible.
%
Finally, the definition of what constitutes a cluster is often overlooked in
most (if not all) works  related to stellar membership estimation.
This is mainly  due to the absence of a standardized definition across different
clustering algorithms,  with each algorithm employing its own
definition.
When applying these algorithms,  selecting cluster members depends
on  several unique parameters  specific to the method used.
 Setting values for these parameters is not  straightforward,
and their choices  often lack a strong justification other than
 producing ``reasonable'' results.

The ideal scenario  would provide a database  containing
both mass and coordinates for each star in full phase space,  encompassing three
positional and three momentum variables. This would allow us to define a
cluster that  properly accounts for the gravitational potential of the Galaxy.
 However, this ideal situation is not the case. We  lack
information  on masses, and even the phase space is  incomplete
 due to the absence of radial velocities in the  vast majority
of cases,\footnote{Currently Gaia contains radial velocity data for less than 2\% of
the observed stars, see: \url{https://www.cosmos.esa.int/web/gaia/dr3}.} What
we have for each observed stars is thus a 5-dimensional data point made up of
coordinates (equatorial, galactic), parallax, and proper motions. With this at
our disposal, we propose the following  general definition of cluster:\\

\noindent\emph{Given an integer value $m>0$ and a point $c$ in an
$n$-dimensional space, a ``cluster'' is defined as the collection of $m$
elements with the smallest $n$-dimensional Euclidean distance to $c$.}\\

\noindent The advantage of explicitly stating a definition of a cluster is that
we no
longer depend on different clustering methods or their extraneous parameters.
Assuming the centre and the estimated number of members are given, this
definition will always yield the same set of selected stars as probable members.
 This is true at least approximately, since uncertainties do play a role
here as we will see below.\\

\begin{figure}
	\includegraphics[width=\columnwidth]{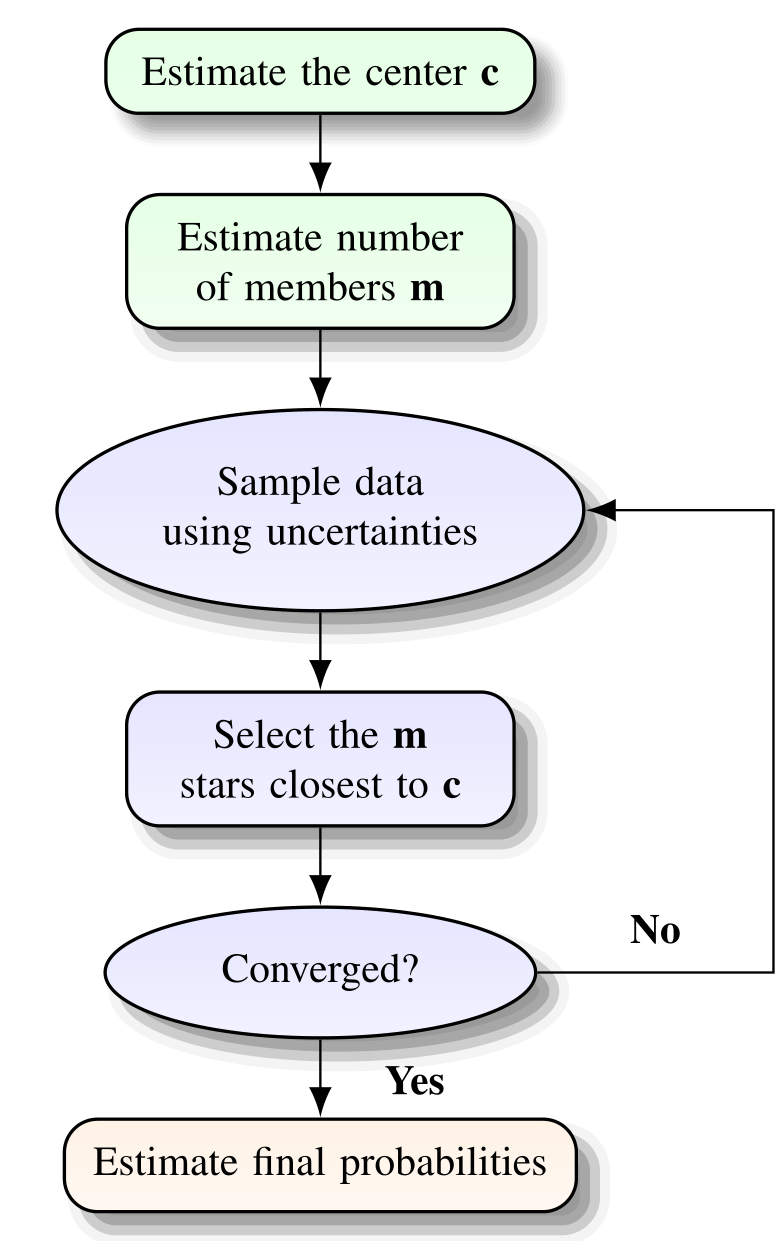}
    \caption{Basic flow chart of the algorithm used by our \texttt{fastMP}
    membership estimation tool.}
    \label{fig:fastMP_code}
\end{figure}

Taking into account the aforementioned shortcomings of general clustering
algorithms, we decided to develop a new tool  for estimating membership
probabilities  which we have named \texttt{fastMP} (acronym for \emph{fast Membership
Probabilities}), based on the  previously outlined definition of a
cluster.
We named the code \texttt{fastMP} due to its processing speed. In our
 extensive testing, it  took \texttt{fastMP} less than 3 seconds
on average to analyse an OC in an 11 years old 4-cores CPU. This means that the
entire UCC (almost 14000 OCs so far) can be processed in approximately 11 hours
on a very modest CPU.  It should be noted that this estimate includes
the time required  for incorporating uncertainties into the process.
In Fig.~\ref{fig:fastMP_code} we  present the  configuration of
the fundamental blocks  within the \texttt{fastMP} algorithm. As can be
seen, it is a rather  straightforward process
  primarily reliant on the two basic parameters mentioned in the
cluster definition: its  central coordinates and the number of members.

We briefly describe each  individual block in the following subsections.
The code is  entirely open source and  distributed under the GNU
General Public License version 3 (GPL
v3),\footnote{\url{https://www.gnu.org/copyleft/gpl.html}} meaning that it
can be easily tested and modified.

\subsection{Centre estimation}

To  estimate the most likely centre $c$ for the OC under analysis, 
\texttt{fastMP} uses a three step process. First, it searches for the region of
maximum density in proper motions. Only the 2-dimensional space of proper
motions is employed here since this is generally where the overdensity is most
visible,  standing out against the surrounding stars in the observed
field. The position of the overdensity is found  through an iterative
process that starts with the full set of data, and gradually ``zooms in'' until
a convergence  criterion is reached.
The second step selects a subset of stars with the closest distance to 
 these centre coordinates in proper motions space. In the third step,
the final centre value is obtained using a k-nearest neighbours algorithm to
select the point with the largest density in the 5-dimensional space 
(coordinates, parallax, proper motions).

\subsection{Number of members estimation}
\label{ssec:members_estim}

Once  the centre point is estimated, the total number $m$ of stars that
can be considered members of the OC is obtained through Ripley's K
function~\citep{ripley_1976,ripley_1979}. This function is used to  assess how
close a group of points is to a random uniform distribution. We refer the
reader to our previous article where we presented pyUPMASK~\citep{Pera_2021},
where we introduced the concept in much more detail. Subsets of stars are
selected in rings, moving outward from the centre values estimated in proper
motions and parallax. If these stars are considered to be far enough from a
random uniform distribution in coordinate space, they are kept as probable
members of the OC. In this block we can also reject stars that are more
likely to belong to other clusters in the frame, to more accurately estimate the
true number of members for the OC under analysis.

Since it is run only once, Ripley's K function can be replaced by any other
method to estimate the size of a cluster (for example some of the already
mentioned clustering algorithms) without much impact on the performance. It can
even be skipped entirely by feeding this number to \texttt{fastMP}, estimated
manually or by some external process, as explained in the final paragraph of
this section.

\subsection{Sampling, selection, convergence}

This is the bootstrap block where we incorporate the data uncertainties into
the final membership probabilities.
Its basic function is to sample the observed data using its uncertainties,
select those $m$ stars closest to the centre $c$, and
finally repeat this process until a convergence criterion is reached. We define
as a stopping condition the run when the total number of stars with probability
greater than 50\% has stabilized for several runs.
The bootstrap block is  usually run a few hundred times before convergence is
achieved.

\subsection{Membership probabilities}

To estimate the membership probability for each star, the algorithm simply
divides the number of times a given star was selected in the bootstrap block by
the number of runs required until convergence.
In  cases where no star was selected by the bootstrap block, probabilities are
assigned based on the 5-dimensional distances to the centre. These are lower
quality probabilities since they do not make use of the uncertainties of the
data.
 This last-resort option is only employed for candidate OCs that are
too faint to be observed, meaning that their members do not form a region
dense enough to distinguish them from the background of field stars.\\

Finally, it is worth noting that an important advantage of \texttt{fastMP} over
tools commonly used like UPMASK, pyUPMASK, HDBSCAN, etc., is that it can run in
supervised mode. In this context, we make the distinction  between
supervised  and unsupervised in the sense that the algorithms mentioned
above work with no prior information about the cluster(s) being analysed. We
call this ``unsupervised''.
In contrast, \texttt{fastMP} allows information about the cluster to be passed
along with the input data. We call this ``supervised''. Such information can
be the centre of the cluster, its total number of members, or both. If any of
these values are fed to the code after estimating them manually or via an
external method, then its corresponding block (either centre or number of
members estimation)is skipped. This is of great help, particularly
for OCs that are very faint or sparse and  cannot easily be picked up by the
usual clustering algorithms.
It is also a feature that allows the code to analyse systems that are very close
together, either in positional space or a combination of this and the proper
motions and/or parallax dimensions, by fixing its centre and/or  the number of
constituent members. In the following section we show that \texttt{fastMP} has
an excellent performance  when compared to recent works that generate lists of
members for OCs, like CANTAT20 and HUNT23.

\section{Results}
\label{sec:results}

This section is divided as follows. In Sect.~\ref{ssec:duplicates} we analyse
the issue of duplicated entries across catalogues. Sect.~\ref{ssec:members}
compares the results of our membership estimation with those from recent
large catalogues. Sect.~\ref{ssec:classif} discusses a possible
classification of the candidate OCs as real physical objects to separate them
from artefacts derived from the application of different clustering algorithms.
Finally, Sect.\ref{ssec:overview} presents a brief overview of the online
service where this catalogue is hosted and a few of the issues that will be
improved in the future.

\subsection{Duplicates}
\label{ssec:duplicates}

One of the main problems with today's state of research in the area of OCs is,
as mentioned earlier, the large number of articles being published on the subject.
This should not be a concern a priori, but with articles appearing every few
months presenting new candidates by the thousands, keeping track of the
latest proposed objects becomes a not so simple task.
This is evidenced by the large number of potential duplications that can be
found when cross-matching the most recent databases with older ones.

In this work we do not attempt to merge and/or discard candidates as duplicates,
as this is not a trivial assessment to make. The UCC only flags OCs that have
the potential of being duplicates of others, following a parallax-based
decision rule. This rule checks the distance from a given OC to all the others
in the catalogue in three separate components: coordinates (arcmin), parallax 
(mas), and proper motions (mas yr$^{-1}$). If these distances are smaller than a
given threshold, they are converted into a probability using a linear relation.
 If they are larger than the threshold, they are assigned a probability
of zero. \footnote{If the distances in all
three components between the OC and another object are zero, then the
probability of these two being duplicates of each other is 1. If all three
distances are beyond the maximum limits shown in the parallax-based rules, then
the probability of duplication is zero. Distance values beyond zero and these
limits are converted linearly in the probability range (0, 1).}
The relations depend on the distance to the OC (estimated
from its catalogued parallax) and can be seen in the block below.

\begin{verbatim}
if parallax >= 4
    xy_r, plx_r, pm_r = 20, 0.5, 1
else 3 <= parallax < 4
    xy_r, plx_r, pm_r = 15, 0.25, 0.75
else 2 <= parallax < 3
    xy_r, plx_r, pm_r = 10, 0.2, 0.5
else 1.5 <= parallax < 2
    xy_r, plx_r, pm_r = 7.5, 0.15, 0.35
else 1 <= parallax < 1.5
    xy_r, plx_r, pm_r = 5, 0.1, 0.25
else .5 <= parallax < 1
    xy_r, plx_r, pm_r = 2.5, 0.075, 0.2
else parallax < .5
    xy_r, plx_r, pm_r = 2, 0.05, 0.15
else parallax < .25
    xy_r, plx_r, pm_r = 1.5, 0.025, 0.1
else parallax is nan
    xy_r, pm_r = 2.5, 0.2
\end{verbatim}

Here, \texttt{xy\_r, plx\_r} and \texttt{pm\_r} are the parallax-based
thresholds for each component (in arcmin, mas, and mas yr$^{-1}$, respectively).
For example, if a candidate OC  named A has a catalogued parallax of
0.75 mas, then \texttt{xy\_r, plx\_r, pm\_r = 2.5, 0.075, 0.2}. This means that 
 if there exists an OC  named B with distances  to
A smaller than those thresholds,  A and B will have a non zero
probability of being duplicates  of the each other.  These rules
for finding possible duplicates translate smaller distances to larger
probabilities.
The reason  for splitting the threshold  into parallax ranges is that the
farther away the OC the smaller its parallax, coordinates radius, and mean
proper motions will tend to be. These thresholds and parallax ranges are of
course entirely arbitrary, but we have observed very reasonable results using
them.
Employing this rule, if the catalogued positions are used, the aforementioned OC
candidate CWDDL 14677 is flagged with a 65\% probability of being a duplicate of
Blanco 1. If we use the values for the main coordinates, proper motions, and
parallax obtained from the most likely members found by \texttt{fastMP} 
 instead, this probability increases to 84\%.\\

\begin{figure}
	\includegraphics[width=\columnwidth]{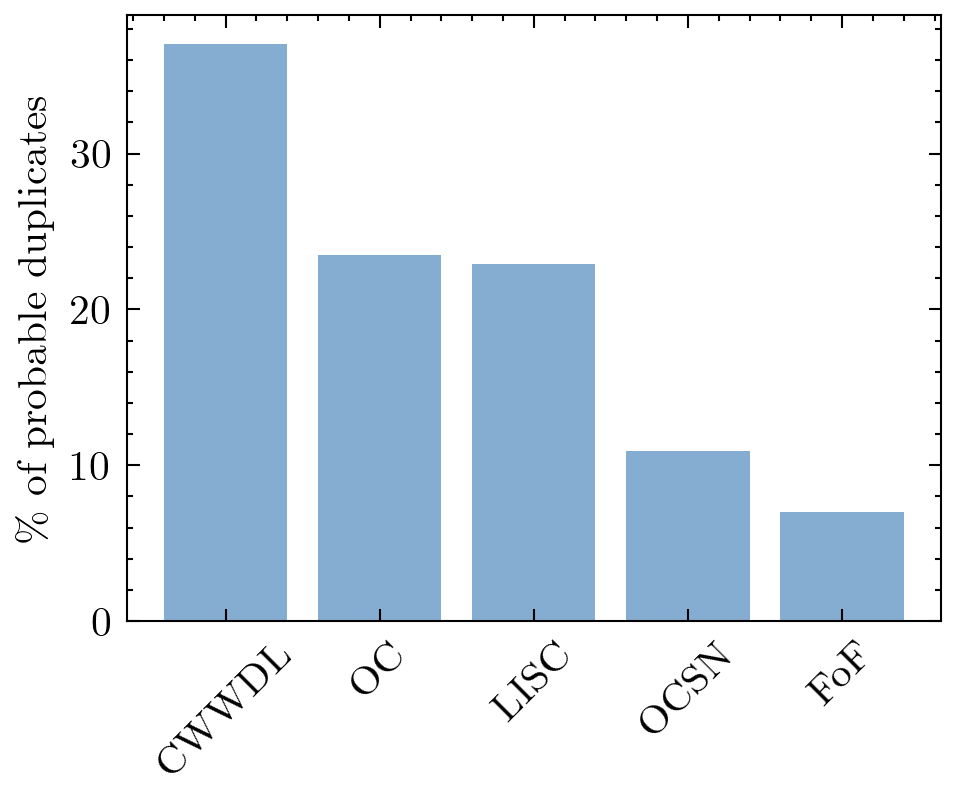}
    \caption{Percentage of probable duplicates for five recent databases of
    candidate OCs presented in the literature, characterized by their IDs.
    See Table~\ref{tab:references} to match the corresponding article(s)
    to each ID.}
    \label{fig:duplicates}
\end{figure}

In Fig.~\ref{fig:duplicates} we show the percentage of entries flagged as
probable duplicates for five of the latest articles mentioned in
Table~\ref{tab:references}. We selected as probable duplicates those entries
with an assigned probability larger than 50\%.
The CWWDL clusters from~\cite{Chi_2023_3} stand out with almost 40\% of its
1179 candidate OCs flagged as probable duplicates of entries in previous (older)
catalogues. This is a rather large value that translates to more than 400 of the
candidate OCs listed in this catalogue.

An example of two candidate OCs flagged as possible duplicates, with a
probability value of 50\% and 90\%, are CWWDL 578~\cite{Chi_2023_3} and
LISC 3279~\citep{Li_2022}. These objects
were associated by our parallax-based rule to two UBC OCs, presented
in~\cite{Castro-Ginard_2020}.
The five dimensional positions catalogued values for these four entries
are shown in Table~\ref{tab:duplicates}.
 While we acknowledge that a simple rule based on distances in the
coordinates, parallax, and proper motion spaces is not a  substitute for a
 thorough and detailed analysis  of duplications, it can
certainly serve as a reasonable starting point. Studies  involving binary
clusters systems can  also rely on these probabilities for their
initial assessment.  In total, there are over 2000
entries in the UUC flagged as  either duplicates or flagged  for
having one or more duplicates.  This accounts for $\sim$15\% of the
entire catalogue.

\begin{table}
	\centering
	\caption{Examples of two pairs of OCs flagged as duplicates by our
	parallax-based rule with probabilities of 50\% (1st and 2nd columns) and
	90\% (3rd and 4th columns).}
	\label{tab:duplicates}
	\begin{tabular}{lcccc}
		\hline
         &  \multicolumn{2}{c}{P$\approx$50\%} & \multicolumn{2}{c}
         {P$\approx$90\%}\\
          &   CWDDL 578 & UBC 395 & LISC 3279  &  UBC 361 \\
		\hline
		 RA &   345.743 & 345.655 & 290.000  &  290.016 \\
		 DEC &   57.231 & 57.206 & 15.146  &  15.157 \\
		 Plx &   0.412 & 0.409 & 0.630  & 0.632 \\
		 pmRA &   -2.844 & -2.869 & -1.669  & -1.701 \\
		 pmDE &   -2.542 & -2.591 & -5.225  & -5.232 \\
		\hline
	\end{tabular}
\end{table}

\subsection{Membership analysis}
\label{ssec:members}

More than one million estimated members are stored in this initial version of
the UCC. This is almost half a million more than those  found in the HUNT23
catalogue (after removing globular cluster and moving groups), and approximately
five times  the number in CANTAT20.  The latter database is expected to
contain  fewer entries not only because it  lists a smaller
number of OCs, but also because it only reaches  a magnitude of G=18
mag, whereas UCC and HUNT23 go two magnitudes  deeper.

\begin{figure}
	\includegraphics[width=\columnwidth]{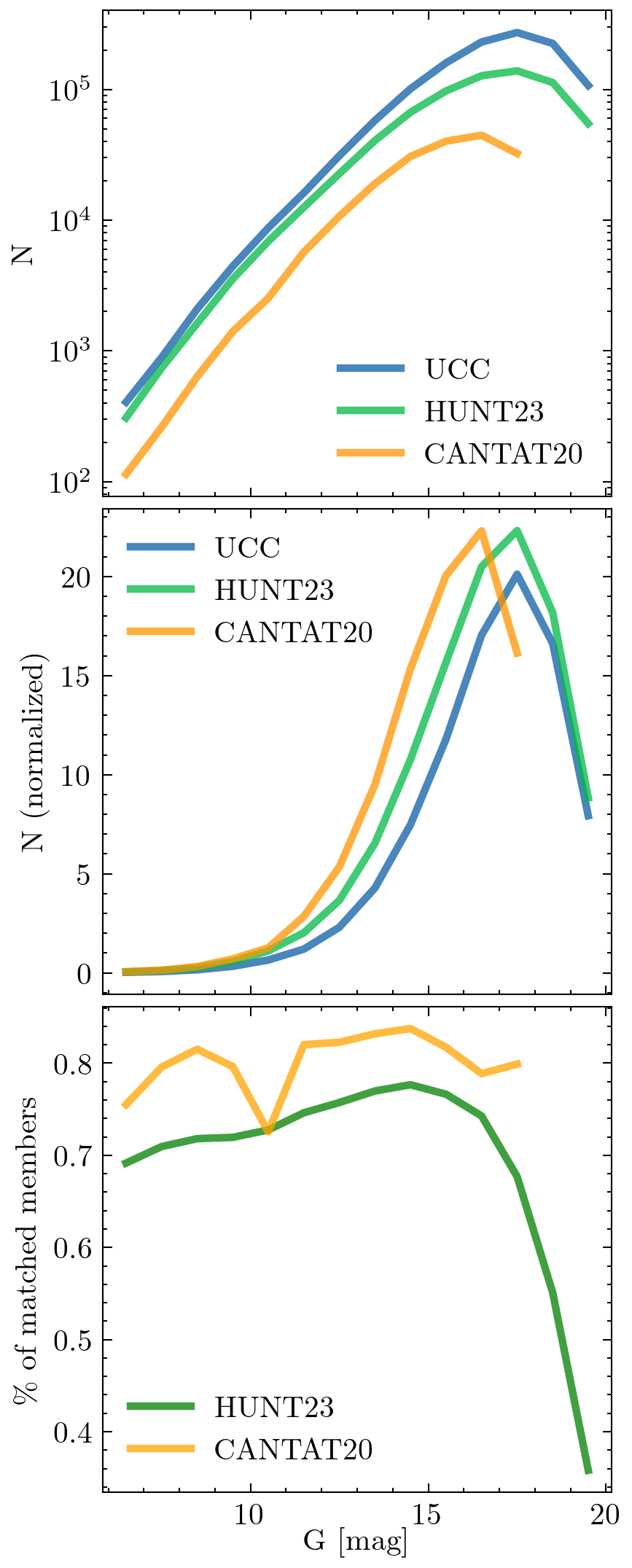}
    \caption{Top: total number of identified members versus magnitude, for
    the catalogues HUNT23, CANTAT20, and the UCC.
     Centre: same as above but normalized by the total number of entries in each
    catalogue.
    Bottom: percentage of members in HUNT23 and CANTAT20 matched with members
    identified by the UCC, versus magnitude.}
    \label{fig:members}
\end{figure}

In the top plot of Fig.~\ref{fig:members}, we show the distribution of
estimated members for these three catalogues  as a function of Gaia's G
magnitude. The  central plot shows the same distributions but normalized
by the total number of entries in each catalogue.
Finally, the bottom plot displays the percentage of
members in CANTAT20 and HUNT23  that match members estimated by the
UCC.  In this plot CANTAT20  shows an overall match in the range
75-80\%  across the entire magnitude  range, up to its
maximum of G=18 mag. The match with HUNT23  remains around 70-75\% up to
G$\approx$17 mag, after which it  starts to decline.
For the largest magnitude in UCC and HUNT23, G=20 mag, the match
percentage is $\sim$35\%. This can be  primarily explained by two processes.
 On one hand, the  large sensitivity of the HDBSCAN algorithm often
 causes it to return false positives, as reported by HUNT23.
This can  result in an overestimation of the number of members assigned to
each candidate OC.
On the other hand, the methods employed in CANTAT20 and HUNT23,  namely
UPMASK and HDBSCAN, respectively, do not  take into account the
uncertainties  in Gaia's data,  while
\texttt{fastMP} does.
Incorporating uncertainties into the membership  probability estimation
 has the most significant impact on stars in the lower mass region, as
 they have the largest errors. The \texttt{fastMP} code also  adopts a
more cautious approach when estimating the total number of stars associated 
 with a given OC, based on Ripley's K function, as mentioned in
Sect.~\ref{ssec:members_estim}.
The  central plot in Fig.~\ref{fig:members}
clearly  illustrates these effects, where normalizing by the number of
OCs in each catalogue  causes the UCC's distribution dip below those of
CANTAT20 and HUNT23.
The UCC  averages $\sim$90 members per OC, HUNT23 has around $\sim$110
members per OC, and CANTAT20 is close to $\sim$95 members per OC.
%

\subsection{Classification}
\label{ssec:classif}

Determining what constitutes a  genuine physical cluster of stars is not
an easy task when dealing with OCs~\citep{Parker_2014}.
Whereas globular clusters  comprise hundreds of thousands of member
stars, OCs are  significantly smaller. In the previous section we
 demonstrated that a reasonable average for the number of members 
 within an OC is $\sim$100 stars, with a  pronounced bias toward
smaller values.
A  comprehensive physical analysis  to ascertain whether a few dozen
stars are gravitationally bound requires information that we currently lack.
 For example, the definition presented in~\cite{Gieles_2011} requires
a reliable estimation of total mass,  age, effective radius, and
 eventually precise velocities.

\begin{figure*}
	\includegraphics[width=\textwidth]{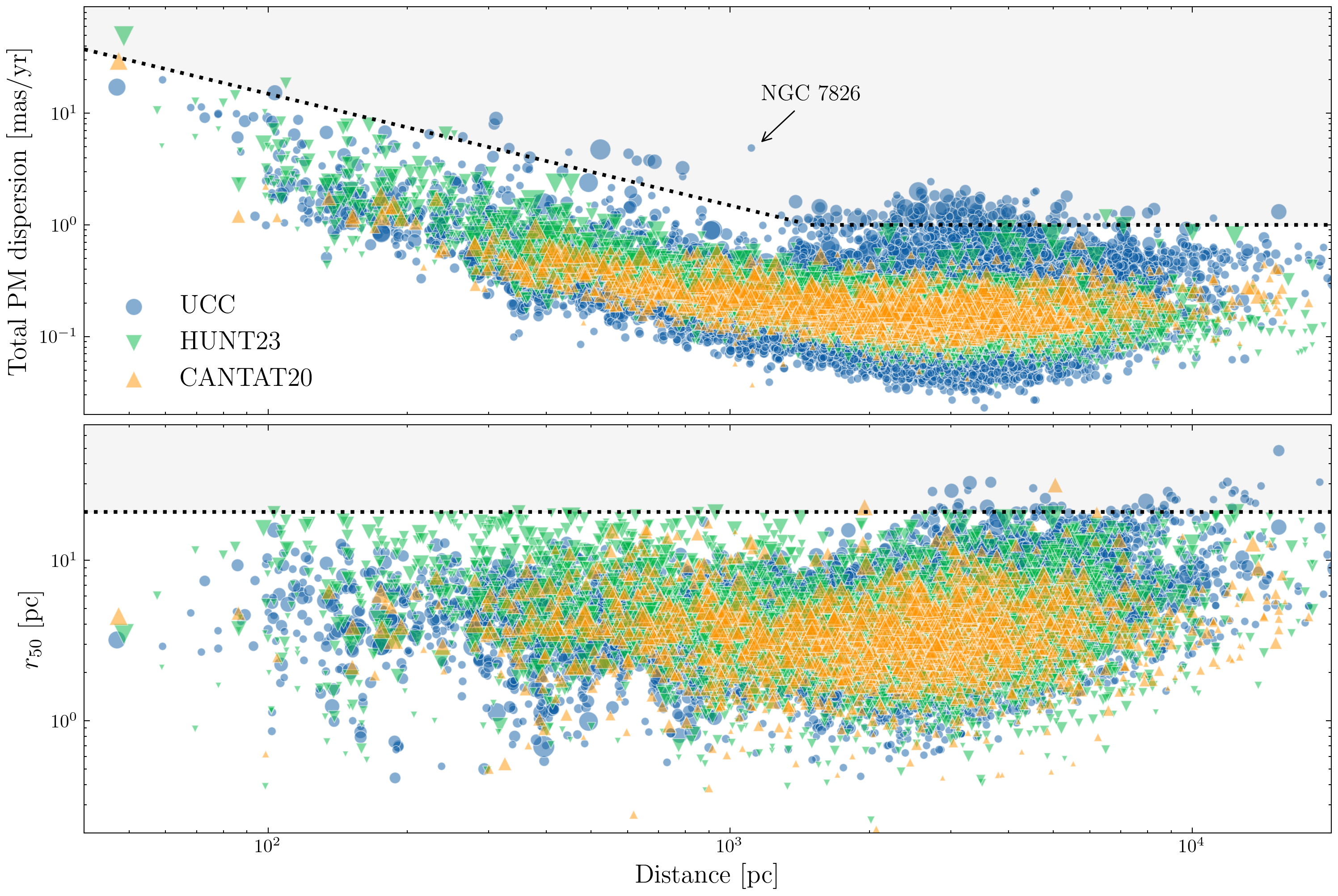}
    \caption{Distribution of total proper motion dispersion (top) and radius
    that contains half the members (bottom) versus distance, for the candidate
    OCs listed in CANTAT20 (orange), HUNT23 (green), and the UCC (blue).
    Sizes are proportional to each candidate's associated number of members.
    The grey region in both plots is the ``not true OC region'' defined by
    the rules proposed by \citet{Cantat-Anders_2020}.}
    \label{fig:pms_plx}
\end{figure*}

This fact notwithstanding, there are still methods we can  employ to
approximate a  comprehensive dynamical analysis  for characterizing
candidate OCs as more or less likely to be  genuine. Two of these
methods, or quality  criteria, were proposed
by~\cite{Cantat-Anders_2020}.
The first  method is based on  setting an upper limit 
 for the internal velocity dispersion of OCs. Beyond  this limit,
objects  are expected to either be globular clusters or unbound groups.
 Initially, this limit is  set conservatively at 5 km\,s$^{-1}$.
 However it is relaxed to 1 mas\,yr$^{-1}$ for candidate OCs 
 located beyond $\sim$1000 pc,  where uncertainties in proper
motions tend to dominate the measured values.
We set the  transition between both  these limits at $\sim$1.5
kpc,  which is the distance  at which the velocity dispersion
lines mentioned  earlier intersect.
The second method  involves measuring the internal spatial dispersion of
an OC.  According to~\cite{Cantat-Anders_2020}, for the majority of
 confirmed OCs, the maximum dimension  containing half the
members  is estimated to  be $\sim$15 pc.  The authors
of HUNT23  slightly relax this condition to 20 pc, which we also adopt 
 here.
In Fig.~\ref{fig:pms_plx} we  display the results of both of these
approaches  for approximating a ``true OC region''  using the
data collected in the UCC,  as well as the data presented in CANTAT20
and HUNT23.
As can be seen, the region occupied  by all three catalogues is very
similar, with the UCC showing a broader distribution in total proper motion
dispersion beyond $\sim$1000 pc.
 Nevertheless, the majority of candidate OCs  fall well below the
proposed  quality criteria lines in both plots.
Objects with large proper motion dispersions are mostly candidates listed only
in the~\cite{Kharchenko_2012} catalogue  which, to the best of our
knowledge,  have never undergone a proper analysis. There are also
$\sim$30 infrared candidates OCs from~\cite{Ryu_2018}  for which
membership estimation is poor.

An example of an object located beyond the quality  line for proper
motion dispersion is NGC 7826,  as identified in the top plot of
Fig.~\ref{fig:pms_plx}.
 While this object  was found to not be a 
 genuine OC in~\cite{Kos_2018} and  classified as an asterism
in~\cite{Cantat-Anders_2020}, we include it  in the UCC as it is
listed  as an OC in the~\cite{Loktin_2017} catalogue.
 As expected, the proper motion distribution of its most likely
associated stars also  places it
 outside the  defined ``true OC region'' in this 
 study.\\

\begin{figure}
	\includegraphics[width=\columnwidth]{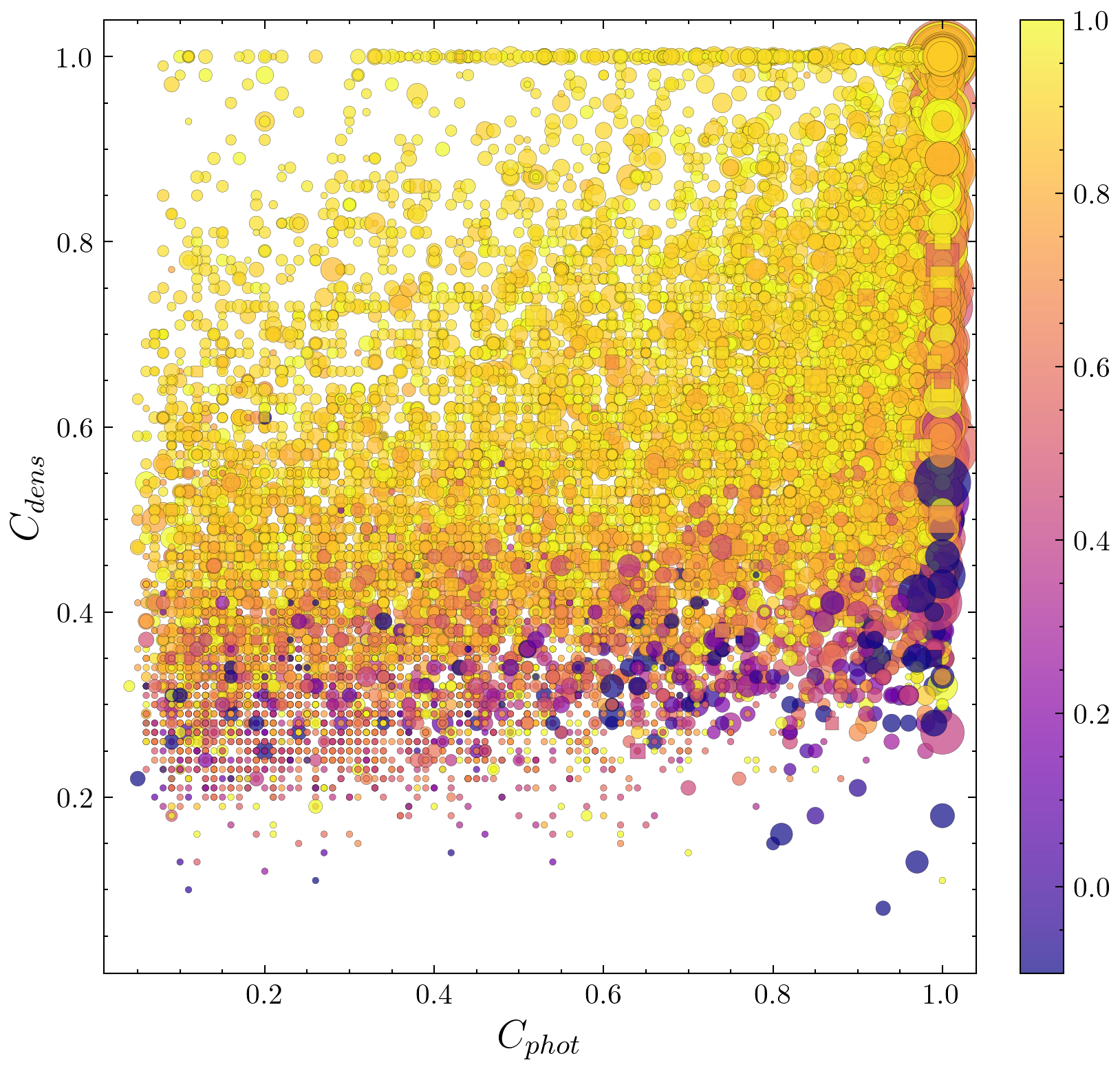}
    \caption{Five dimensional density $C_{dens}$ versus photometry based metric
    $C_{phot}$ for the candidate OCs listed in the UCC. Colour is related to the
    vertical distance to the total proper motion quality cut. The colourbar is
    clipped at -0.1 and 1 to improve visibility. Size is related to
    the spatial extension of its members. Squares are
    associated to candidates beyond the 20 pc spatial dispersion quality cut.}
    \label{fig:classification}
\end{figure}

We can also define other metrics to classify candidate OCs  as more or
less likely  to be true physical objects,  based on their
estimated members' data.
The first metric we developed is a density-based classification, 
 denoted as $C_{dens}$, and the second one is a photometry-based
classification,  denoted as $C_{phot}$.
These  metrics bear conceptual similarities to the CST score and CMD
class defined in HUNT23, respectively.
The density-based metric compares the distribution of member stars to 
 that of nearby field stars in the 5-dimensional space of coordinates,
proper motions, and parallax. The reasonable expectation is that neighbouring
cluster members should  exhibit smaller average distances than
neighbouring field stars.
 Photometric data is processed separately because member stars of an OC in a
colour-magnitude diagram (CMD) do not cluster around a central value, as
they do in  other data dimensions. Instead, they are distributed
 along an elongated path across the evolutionary sequence. For this
reason we do not employ a closest-neighbour density-based method, but one based
on the likelihood of the members' sequence being equivalent to a random
sequence drawn from field stars. To quantify this we  utilize the same
function  developed for our \texttt{ASteCA} package~\citep{Perren_2015},
 which is based on the Poissonian distribution likelihood defined
in~\cite{Tremmel_2013}.
In Fig.~\ref{fig:classification}, the distribution of these two metrics 
 is shown.  They are both normalized  within the [0, 1]
range, where 1  indicates a higher likelihood of being a collection of
related cluster member stars  for either metric. The colour is assigned
 based on the vertical distance to the quality  line in the
total proper motion dispersion diagram shown in the top plot of
Fig.~\ref{fig:pms_plx}.
Objects that  fall beyond this limit (i.e.,  displaying larger
proper motion dispersion than that allowed for an OC) have values below zero and
are drawn in purple.  The sizes of the markers correspond to the spatial
extension of the estimated members.
A clear  trend is evident where candidate OCs with larger $C_{dens}$
values  tend to have large $C_{phot}$ values,  as expected.
There is also a visible  dispersion around the 1:1 identity relation,
 indicating these two metrics are not entirely correlated. This 
 lack of correlation is desirable, as it ensures that both methods
provide distinct information. Objects with large total proper motion
dispersions are mostly  associated with low $C_{dens}$ values but tend
to span the entire range of $C_{phot}$ values.

\begin{figure}
	\includegraphics[width=\columnwidth]{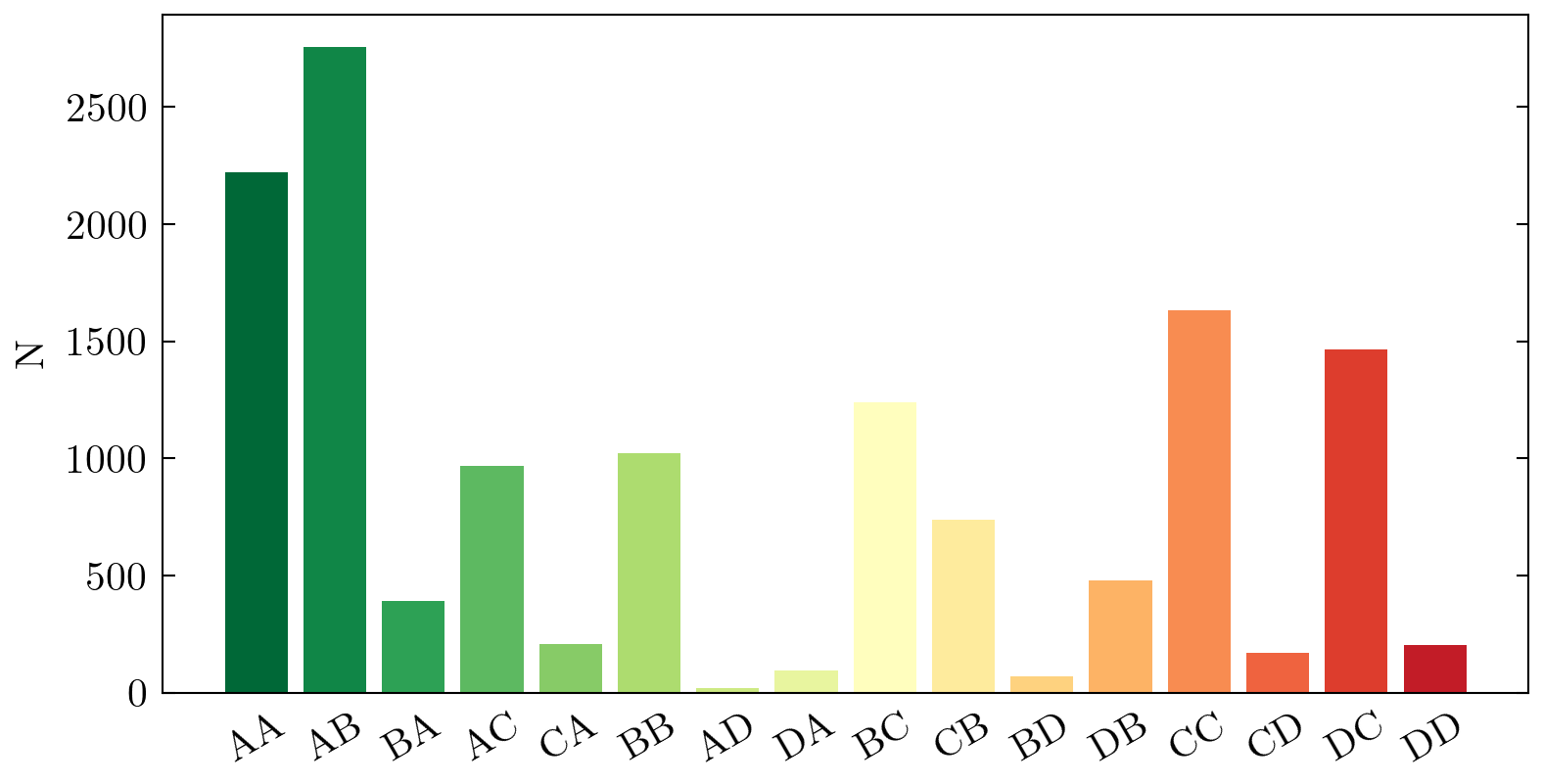}
    \caption{Distribution of the quality classes obtained combining the
    $C_{phot}$ and $C_{dens}$ values for each candidate OC as described in the
    text.}
    \label{fig:classif_bar}
\end{figure}

We  combine these two classification metrics into a single quality
class for each candidate OC to provide a quick  overview of the
characteristics of the catalogued objects.
First, we  divide the [0, 1] range for both metrics into four
equal-length segments (from 0 to 0.25, from 0.25 to 0.5, etc.) and assign a
letter to each  segment.  The assignments range from D for
the [0, 0.25] segment to A for the [0.75, 1] segment.
These two letters, one for each metric, are  then combined to generate a
single class
out of 16 possible combinations. The letter  corresponding to the
$C_{phot}$ value is  placed first, followed by the letter obtained for
the $C_{dens}$ value for  each object. The better quality clusters are
thus assigned AA classes  while lower quality ones  receive
DD classes.
The complete distribution of  these classes is shown in
Fig.~\ref{fig:classif_bar}.
The three  highest quality classes (AA, AB, and BA)
 encompass over 5300 objects, constituting $\sim$40\% of the catalogue.
 In contrast the three lowest classes (CD, DC, and DD) contain 
 nearly 2000 candidate OCs,  equivalent to $\sim$15\% of the
UCC. Classes AD and DA are  among the least populated, 
 suggesting that it is unlikely to  encounter an object with a
large value in $C_{phot}$ and a low value in $C_{dens}$ or vice versa, as one
would expect.\\

\begin{figure*}
	\includegraphics[width=\textwidth]{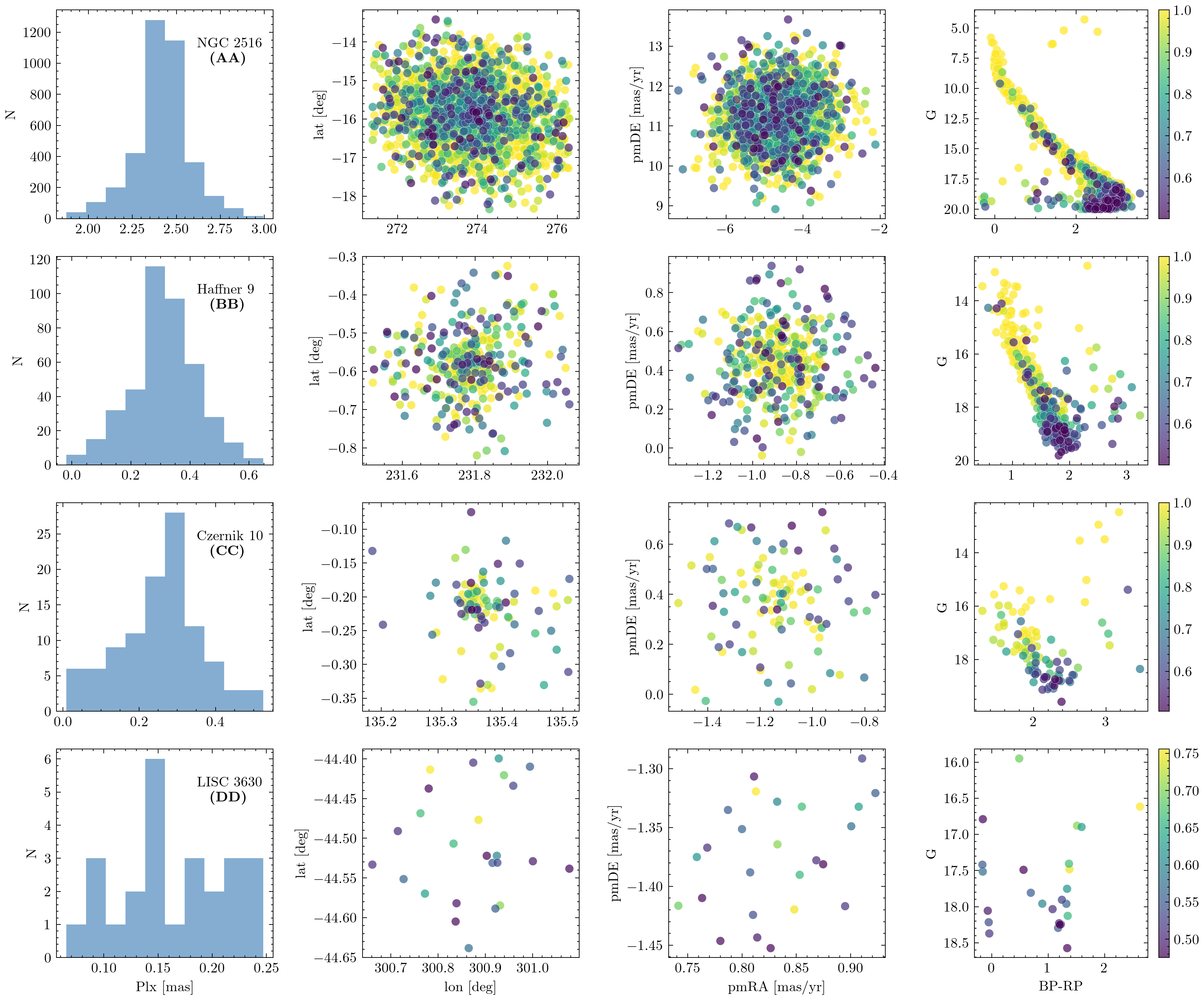}
    \caption{Examples of  candidate OCs with combined classes AA, BB,
    CC, and DD from top to bottom rows, respectively. Plots show from left to right:
    parallax distribution, spatial dispersion, proper motions, and CMD.
    Colourbars are associated to the membership probabilities assigned by 
    \texttt{fastMP}.}
    \label{fig:classif_examples}
\end{figure*}

In Fig.~\ref{fig:classif_examples} we  present examples of four OCs
listed in the UCC,  each assigned combined classes of AA, BB, CC, and
DD. The  distinction between the better AA and  the lower-quality
DD classes is evident, both in the  denser spatial, proper
motions, and parallax distributions, as well as  in the  more defined
cluster sequence  observed in the CMD.
 It is important to note however that these quality cuts and
classification methods,  while valuable, are not sufficient to entirely
replace a proper analysis of a candidate OC.
 They serve as guides to identify problematic cases, but their 
 effectiveness still  relies heavily on the precision of
the membership estimation process.
 In cases where this process performs poorly, especially for
more  complex cases,  the results  may be compromised.
 Improved data quality for observed stars in the near future
 will hopefully enhance the effectiveness of these methods.

\subsection{Overview of the service}
\label{ssec:overview}

To facilitate cross-catalogues identification of entries, each candidate OC in
the UCC is assigned a unique name following the  specifications of
the International Astronomical Union. The naming convention is
formatted as \emph{UCCGLLL.LsBB.B}, where \emph{UCC}  represents the
catalogue's name, \emph{G} indicates  the use of galactic coordinates, 
\emph{LLL.L} is the (truncated) longitude
for the object, \emph{s} is the sign of its latitude, and \emph{BB.B} is the 
(truncated) latitude.  In cases where two objects share 
 the same coordinates, a lowercase letter from
a to z is  appended to the end of the name.

The spatial position of each entry in the catalogue is displayed  using
Aladin's visualization tool,\footnote{Aladin: 
\url{http://aladin.cds.unistra.fr/}}
 showcasing coordinates, proper motions, parallax, and radial velocity
values when  available in the literature.
Links to search the cluster's main name in the SAO/NASA Astrophysics Data System
(ADS),\footnote{SAO/NASA ADS: \url{https://ui.adsabs.harvard.edu/}} as well as
 links to perform a region search in the Strasbourg astronomical Data
Center (CDS)\footnote{CDS: \url{http://cdsportal.u-strasbg.fr/}} are provided.
A Python notebook hosted  on Google's Colaboratory
service\footnote{Colaboratory: \url{https://colab.research.google.com/}} is made
available,  enabling users to interactively explore the Gaia survey data
of each candidate's estimated members. In this initial version the UCC only
contains membership data obtained through our \texttt{fastMP} tool. 
 However, in future updates we  plan to include members
estimated by other works, such as those from CANTAT20 and HUNT23, as well as the
most recent catalogues.

Fundamental parameters  like distance, extinction, age, and metallicity
are also  presented when available,  sourced from various
databases. These  details will be expanded  upon in future
UCC updates as more values from the literature are incorporated.
For each candidate OC nearby entries and  potential duplicates are
 displayed, along with the probability of being a duplicate, as
 explained in Sect.~\ref{ssec:duplicates}.

In an upcoming version of the UCC we will  include individual notes
 on OCs, whenever available.  Obtaining these notes is challenging as
they are scattered throughout the literature  rather than compiled in
databases.
For example~\cite{Cantat-Anders_2020}  provides in its appendix a list
of OCs classified as asterisms,  valuable information  that is
lost if only the candidate's parameters are displayed in the catalogue.


\section{Conclusions}
\label{sec:conclusions}

We  introduced the Unified Cluster Catalogue or UCC, along with its
accompanying tool for membership probability estimation, the \texttt{fastMP}
code. This is catalogue  represents the most extensive compilation  of
open clusters to date and will be regularly updated as new databases 
 become publicly available in the  scientific literature.
The UCC is accessible through its dedicated website at \url{https://ucc.ar},
where each OC is  presented along with its fundamental parameters,
 sourced from the literature  whenever available.  Users
can interactively explore the data online through Python notebooks hosted 
 on the Google Colaboratory service. In its
initial version the UCC  includes nearly 14000 unique candidate OCs,
 encompassing a combined total of over 1 million proposed member stars,
averaging approximately $\sim$90 member stars per OC.

Replacing the trained eye of a researcher even for the initial assessment
of what constitutes a true OC in a completely generalized approach, is
 a formidable challenge. With hundreds or even thousands of
new candidate OCs  emerging in the literature every few months,
 the need for a systematic method  to address this issue becomes
 increasingly crucial.
Our classification parameters, $C_{phot}$ and $C_{dens}$, were 
 developed to assist in this task, but a  thorough
visual inspection  remains essential, particularly for more 
 complex objects.
We  anticipate that the Unified Cluster Catalogue and its associated
online service  will prove to be a valuable resource for the
astrophysical research community.  We welcome all suggestions 
 for expanding and improving it in the future.

\section*{Acknowledgements}

 We thank the financial support from the University of La Plata
(I+D 11/G148), as well as the financial support from the University of Rosario 
(PPCT 80020210300042UR). We also thank the financial support from the IALP 
(CONICET-UNLP).
The authors would like to thank Dr Emily Hunt for her assistance with the
processing of the HUNT23 database.
This work has made use of data from the European Space Agency (ESA) mission
{\it Gaia} (\url{https://www.cosmos.esa.int/gaia}), processed by the {\it Gaia}
Data Processing and Analysis Consortium (DPAC,
\url{https://www.cosmos.esa.int/web/gaia/dpac/consortium}). Funding for the DPAC
has been provided by national institutions, in particular the institutions
participating in the {\it Gaia} Multilateral Agreement.
This research has made use of the WEBDA database, operated at the Department of
Theoretical Physics and Astrophysics of the Masaryk University.
This research has made use of the VizieR catalog access tool, operated at CDS,
Strasbourg, France~\citep{Ochsenbein_2000}.
This research has made use of ``Aladin sky atlas'' developed at
CDS, Strasbourg Observatory, France~\citep{Bonnarel2000,Boch2014,Baumann_2022}.
This research has made use of NASA's Astrophysics Data System.
This research made use of the Python language~\citep{vanRossum_1995}
and the following packages:
NumPy\footnote{\url{http://www.numpy.org/}}~\citep{vanDerWalt_2011},
SciPy\footnote{\url{http://www.scipy.org/}}~\citep{Jones_2001},
matplotlib\footnote{\url{http://matplotlib.org/}}~\citep{hunter_2007},
scikit-learn\footnote{\url{https://scikit-learn.org/}}~\citep{scikit-learn},
\texttt{ASteCA}\footnote{\url{https://github.com/asteca}}~\citep{Perren_2015}.
This work made use of Astropy:\footnote{\url{http://www.astropy.org}} a
community-developed core Python package and an ecosystem of tools and resources
for astronomy \citep{astropy:2013, astropy:2018, astropy:2022}.

\section*{Data Availability}
 
The data underlying this article are available in the repositories associated to
the Unified Cluster Catalogue, accessible at \url{https://github.com/ucc23}.
The code employed to process the data and generate the images in this article
con be found in the repositories for the \texttt{fastMP} code at 
\url{https://github.com/Gabriel-p/fastMP}, and in the repository for the article
itself at \url{https://github.com/gabriel-p-artcls/23-08_UCC}. Any missing
data file and/or code file can be requested to the corresponding author and we
will gladly make it available.



\bibliographystyle{mnras}
\bibliography{biblio} 

\begin{thebibliography}{}
\makeatletter
\relax
\def\mn@urlcharsother{\let\do\@makeother \do\$\do\&\do\#\do\^\do\_\do\%\do\~}
\def\mn@doi{\begingroup\mn@urlcharsother \@ifnextchar [ {\mn@doi@}
  {\mn@doi@[]}}
\def\mn@doi@[#1]#2{\def\@tempa{#1}\ifx\@tempa\@empty \href
  {http://dx.doi.org/#2} {doi:#2}\else \href {http://dx.doi.org/#2} {#1}\fi
  \endgroup}
\def\mn@eprint#1#2{\mn@eprint@#1:#2::\@nil}
\def\mn@eprint@arXiv#1{\href {http://arxiv.org/abs/#1} {{\tt arXiv:#1}}}
\def\mn@eprint@dblp#1{\href {http://dblp.uni-trier.de/rec/bibtex/#1.xml}
  {dblp:#1}}
\def\mn@eprint@#1:#2:#3:#4\@nil{\def\@tempa {#1}\def\@tempb {#2}\def\@tempc
  {#3}\ifx \@tempc \@empty \let \@tempc \@tempb \let \@tempb \@tempa \fi \ifx
  \@tempb \@empty \def\@tempb {arXiv}\fi \@ifundefined
  {mn@eprint@\@tempb}{\@tempb:\@tempc}{\expandafter \expandafter \csname
  mn@eprint@\@tempb\endcsname \expandafter{\@tempc}}}

\bibitem[\protect\citeauthoryear{Anders, Castro-Ginard, Casado, Jordi  \&
  Balaguer-N{\'{u}}{\~{n}}ez}{Anders et~al.}{2022}]{Anders_2022}
Anders F.,  Castro-Ginard A.,  Casado J.,  Jordi C.,
  Balaguer-N{\'{u}}{\~{n}}ez L.,  2022, \mn@doi [Research Notes of the {AAS}]
  {10.3847/2515-5172/ac6034}, 6, 58

\bibitem[\protect\citeauthoryear{{Astropy Collaboration} et~al.,}{{Astropy
  Collaboration} et~al.}{2013}]{astropy:2013}
{Astropy Collaboration} et~al., 2013, \mn@doi [\aap]
  {10.1051/0004-6361/201322068}, \href
  {http://adsabs.harvard.edu/abs/2013A%26A...558A..33A} {558, A33}

\bibitem[\protect\citeauthoryear{{Astropy Collaboration} et~al.,}{{Astropy
  Collaboration} et~al.}{2018}]{astropy:2018}
{Astropy Collaboration} et~al., 2018, \mn@doi [\aj] {10.3847/1538-3881/aabc4f},
  \href {https://ui.adsabs.harvard.edu/abs/2018AJ....156..123A} {156, 123}

\bibitem[\protect\citeauthoryear{{Astropy Collaboration} et~al.,}{{Astropy
  Collaboration} et~al.}{2022}]{astropy:2022}
{Astropy Collaboration} et~al., 2022, \mn@doi [apj] {10.3847/1538-4357/ac7c74},
  \href {https://ui.adsabs.harvard.edu/abs/2022ApJ...935..167A} {935, 167}

\bibitem[\protect\citeauthoryear{{Babusiaux} et~al.,}{{Babusiaux}
  et~al.}{2022}]{Babusiaux_2022}
{Babusiaux} C.,  et~al., 2022, \mn@doi [arXiv e-prints]
  {10.48550/arXiv.2206.05989}, \href
  {https://ui.adsabs.harvard.edu/abs/2022arXiv220605989B} {p. arXiv:2206.05989}

\bibitem[\protect\citeauthoryear{{Barb{\'a}} et~al.,}{{Barb{\'a}}
  et~al.}{2015}]{Barba_2015}
{Barb{\'a}} R.~H.,  et~al., 2015, \mn@doi [\aap] {10.1051/0004-6361/201424048},
  \href {https://ui.adsabs.harvard.edu/abs/2015A&A...581A.120B} {581, A120}

\bibitem[\protect\citeauthoryear{{Bastian}}{{Bastian}}{2019}]{Bastian_2019}
{Bastian} U.,  2019, \mn@doi [\aap] {10.1051/0004-6361/201936595}, \href
  {https://ui.adsabs.harvard.edu/abs/Bastian_2019} {630, L8}

\bibitem[\protect\citeauthoryear{{Baumann}, {Boch}, {Pineau}, {Fernique}, {Bot}
   \& {Allen}}{{Baumann} et~al.}{2022}]{Baumann_2022}
{Baumann} M.,  {Boch} T.,  {Pineau} F.-X.,  {Fernique} P.,  {Bot} C.,   {Allen}
  M.,  2022, in {Ruiz} J.~E.,  {Pierfedereci} F.,   {Teuben} P.,  eds,
  Astronomical Society of the Pacific Conference Series Vol. 532, Astronomical
  Society of the Pacific Conference Series. p.~7

\bibitem[\protect\citeauthoryear{{Bica}, {Pavani}, {Bonatto}  \& {Lima}}{{Bica}
  et~al.}{2019}]{Bica_2019}
{Bica} E.,  {Pavani} D.~B.,  {Bonatto} C.~J.,   {Lima} E.~F.,  2019, \mn@doi
  [\aj] {10.3847/1538-3881/aaef8d}, \href
  {https://ui.adsabs.harvard.edu/abs/2019AJ....157...12B} {157, 12}

\bibitem[\protect\citeauthoryear{{Boch} \& {Fernique}}{{Boch} \&
  {Fernique}}{2014}]{Boch2014}
{Boch} T.,  {Fernique} P.,  2014, in {Manset} N.,  {Forshay} P.,  eds,
  Astronomical Society of the Pacific Conference Series Vol. 485, Astronomical
  Data Analysis Software and Systems XXIII. p.~277

\bibitem[\protect\citeauthoryear{{Bonatto}, {Kerber}, {Bica}  \&
  {Santiago}}{{Bonatto} et~al.}{2006}]{Bonatto_2006}
{Bonatto} C.,  {Kerber} L.~O.,  {Bica} E.,   {Santiago} B.~X.,  2006, \mn@doi
  [A\&A] {10.1051/0004-6361:20053573}, \href
  {https://ui.adsabs.harvard.edu/abs/2006A&A...446..121B} {446, 121}

\bibitem[\protect\citeauthoryear{{Bonnarel} et~al.,}{{Bonnarel}
  et~al.}{2000}]{Bonnarel2000}
{Bonnarel} F.,  et~al., 2000, \mn@doi [AAPS] {10.1051/aas:2000331}, \href
  {http://cdsads.u-strasbg.fr/abs/2000A%26AS..143...33B} {143, 33}

\bibitem[\protect\citeauthoryear{Campello, Moulavi  \& Sander}{Campello
  et~al.}{2013}]{Campello_2013}
Campello R. J. G.~B.,  Moulavi D.,   Sander J.,  2013, in Pei J.,  Tseng V.~S.,
   Cao L.,  Motoda H.,   Xu G.,  eds, Advances in Knowledge Discovery and Data
  Mining. Springer Berlin Heidelberg, Berlin, Heidelberg, pp 160--172

\bibitem[\protect\citeauthoryear{{Cantat-Gaudin} \& {Anders}}{{Cantat-Gaudin}
  \& {Anders}}{2020}]{Cantat-Anders_2020}
{Cantat-Gaudin} T.,  {Anders} F.,  2020, \mn@doi [\aap]
  {10.1051/0004-6361/201936691}, \href
  {https://ui.adsabs.harvard.edu/abs/2020A&A...633A..99C} {633, A99}

\bibitem[\protect\citeauthoryear{{Cantat-Gaudin} et~al.,}{{Cantat-Gaudin}
  et~al.}{2020}]{Cantat-Gaudin_2020}
{Cantat-Gaudin} T.,  et~al., 2020, \mn@doi [\aap]
  {10.1051/0004-6361/202038192}, \href
  {https://ui.adsabs.harvard.edu/abs/2020A&amp;A...640A...1C} {640, A1}

\bibitem[\protect\citeauthoryear{{Casado}}{{Casado}}{2021}]{Casado_2021}
{Casado} J.,  2021, \mn@doi [Research in Astronomy and Astrophysics]
  {10.1088/1674-4527/21/5/117}, \href
  {https://ui.adsabs.harvard.edu/abs/2021RAA....21..117C} {21, 117}

\bibitem[\protect\citeauthoryear{Casado \& Hendy}{Casado \&
  Hendy}{2023}]{Casado_2023}
Casado J.,  Hendy Y.,  2023, \mn@doi [Monthly Notices of the Royal Astronomical
  Society] {10.1093/mnras/stad071}, 521, 1399

\bibitem[\protect\citeauthoryear{{Castro-Ginard}, {Jordi}, {Luri}, {Julbe},
  {Morvan}, {Balaguer-N{\'u}{\~n}ez}  \& {Cantat-Gaudin}}{{Castro-Ginard}
  et~al.}{2018}]{Castro_Ginard_2018}
{Castro-Ginard} A.,  {Jordi} C.,  {Luri} X.,  {Julbe} F.,  {Morvan} M.,
  {Balaguer-N{\'u}{\~n}ez} L.,   {Cantat-Gaudin} T.,  2018, \mn@doi [\aap]
  {10.1051/0004-6361/201833390}, \href
  {https://ui.adsabs.harvard.edu/abs/2018A&amp;A...618A..59C} {618, A59}

\bibitem[\protect\citeauthoryear{{Castro-Ginard}, {Jordi}, {Luri},
  {Cantat-Gaudin}  \& {Balaguer-N{\'u}{\~n}ez}}{{Castro-Ginard}
  et~al.}{2019}]{Castro-Ginard_2019}
{Castro-Ginard} A.,  {Jordi} C.,  {Luri} X.,  {Cantat-Gaudin} T.,
  {Balaguer-N{\'u}{\~n}ez} L.,  2019, \mn@doi [\aap]
  {10.1051/0004-6361/201935531}, \href
  {https://ui.adsabs.harvard.edu/abs/2019A&amp;A...627A..35C} {627, A35}

\bibitem[\protect\citeauthoryear{{Castro-Ginard} et~al.,}{{Castro-Ginard}
  et~al.}{2020}]{Castro-Ginard_2020}
{Castro-Ginard} A.,  et~al., 2020, \mn@doi [\aap]
  {10.1051/0004-6361/201937386}, \href
  {https://ui.adsabs.harvard.edu/abs/2020A&amp;A...635A..45C} {635, A45}

\bibitem[\protect\citeauthoryear{{Castro-Ginard} et~al.,}{{Castro-Ginard}
  et~al.}{2022}]{Castro-Ginard_2022}
{Castro-Ginard} A.,  et~al., 2022, \mn@doi [\aap]
  {10.1051/0004-6361/202142568}, \href
  {https://ui.adsabs.harvard.edu/abs/2022A&amp;A...661A.118C} {661, A118}

\bibitem[\protect\citeauthoryear{{Chi}, {Wang}  \& {Li}}{{Chi}
  et~al.}{2023a}]{Chi_2023}
{Chi} H.,  {Wang} F.,   {Li} Z.,  2023a, \mn@doi [Research in Astronomy and
  Astrophysics] {10.1088/1674-4527/accbad}, \href
  {https://ui.adsabs.harvard.edu/abs/2023RAA....23f5008C} {23, 065008}

\bibitem[\protect\citeauthoryear{{Chi}, {Wei}, {Wang}  \& {Li}}{{Chi}
  et~al.}{2023b}]{Chi_2023_2}
{Chi} H.,  {Wei} S.,  {Wang} F.,   {Li} Z.,  2023b, \mn@doi [\apjs]
  {10.3847/1538-4365/acb2cc}, \href
  {https://ui.adsabs.harvard.edu/abs/2023ApJS..265...20C} {265, 20}

\bibitem[\protect\citeauthoryear{{Chi}, {Wang}, {Wang}, {Deng}  \& {Li}}{{Chi}
  et~al.}{2023c}]{Chi_2023_3}
{Chi} H.,  {Wang} F.,  {Wang} W.,  {Deng} H.,   {Li} Z.,  2023c, \mn@doi
  [\apjs] {10.3847/1538-4365/accb50}, \href
  {https://ui.adsabs.harvard.edu/abs/2023ApJS..266...36C} {266, 36}

\bibitem[\protect\citeauthoryear{{Dias}, {Monteiro}, {Moitinho}, {L{\'e}pine},
  {Carraro}, {Paunzen}, {Alessi}  \& {Villela}}{{Dias}
  et~al.}{2021}]{Dias_2021}
{Dias} W.~S.,  {Monteiro} H.,  {Moitinho} A.,  {L{\'e}pine} J.~R.~D.,
  {Carraro} G.,  {Paunzen} E.,  {Alessi} B.,   {Villela} L.,  2021, \mn@doi
  [\mnras] {10.1093/mnras/stab770}, \href
  {https://ui.adsabs.harvard.edu/abs/2021MNRAS.504..356D} {504, 356}

\bibitem[\protect\citeauthoryear{{Dreyer}}{{Dreyer}}{1888}]{Dreyer1888}
{Dreyer} J.~L.~E.,  1888, \memras, \href
  {https://ui.adsabs.harvard.edu/abs/1888MmRAS..49....1D} {49, 1}

\bibitem[\protect\citeauthoryear{Efron}{Efron}{1979}]{Efron_1979}
Efron B.,  1979, \mn@doi [The Annals of Statistics] {10.1214/aos/1176344552},
  7, 1

\bibitem[\protect\citeauthoryear{Ester, Kriegel, Sander  \& Xu}{Ester
  et~al.}{1996}]{Ester_1996}
Ester M.,  Kriegel H.-P.,  Sander J.,   Xu X.,  1996, in Knowledge Discovery
  and Data Mining.

\bibitem[\protect\citeauthoryear{{Ferreira}, {Santos}, {Corradi}, {Maia}  \&
  {Angelo}}{{Ferreira} et~al.}{2019}]{Ferreira_2019}
{Ferreira} F.~A.,  {Santos} J.~F.~C.,  {Corradi} W.~J.~B.,  {Maia} F.~F.~S.,
  {Angelo} M.~S.,  2019, \mn@doi [\mnras] {10.1093/mnras/sty3511}, \href
  {https://ui.adsabs.harvard.edu/abs/2019MNRAS.483.5508F} {483, 5508}

\bibitem[\protect\citeauthoryear{{Ferreira}, {Corradi}, {Maia}, {Angelo}  \&
  {Santos}}{{Ferreira} et~al.}{2020}]{Ferreira_2020}
{Ferreira} F.~A.,  {Corradi} W.~J.~B.,  {Maia} F.~F.~S.,  {Angelo} M.~S.,
  {Santos} J.~F.~C. J.,  2020, \mn@doi [\mnras] {10.1093/mnras/staa1684}, \href
  {https://ui.adsabs.harvard.edu/abs/2020MNRAS.496.2021F} {496, 2021}

\bibitem[\protect\citeauthoryear{{Ferreira}, {Corradi}, {Maia}, {Angelo}  \&
  {Santos}}{{Ferreira} et~al.}{2021}]{Ferreira_2021}
{Ferreira} F.~A.,  {Corradi} W.~J.~B.,  {Maia} F.~F.~S.,  {Angelo} M.~S.,
  {Santos} J.~F.~C. J.,  2021, \mn@doi [\mnras] {10.1093/mnrasl/slab011}, \href
  {https://ui.adsabs.harvard.edu/abs/2021MNRAS.502L..90F} {502, L90}

\bibitem[\protect\citeauthoryear{{Friel}}{{Friel}}{1995}]{Friel1995}
{Friel} E.~D.,  1995, \mn@doi [\araa] {10.1146/annurev.aa.33.090195.002121},
  \href {https://ui.adsabs.harvard.edu/abs/1995ARA&A..33..381F} {33, 381}

\bibitem[\protect\citeauthoryear{{Froebrich}, {Scholz}  \&
  {Raftery}}{{Froebrich} et~al.}{2007}]{Froebrich_2007}
{Froebrich} D.,  {Scholz} A.,   {Raftery} C.~L.,  2007, \mn@doi [\mnras]
  {10.1111/j.1365-2966.2006.11148.x}, \href
  {https://ui.adsabs.harvard.edu/abs/2007MNRAS.374..399F} {374, 399}

\bibitem[\protect\citeauthoryear{{Gaia Collaboration} et~al.,}{{Gaia
  Collaboration} et~al.}{2016}]{Gaia_2016}
{Gaia Collaboration} et~al., 2016, \mn@doi [\aap]
  {10.1051/0004-6361/201629272}, \href
  {https://ui.adsabs.harvard.edu/abs/2016A&A...595A...1G} {595, A1}

\bibitem[\protect\citeauthoryear{{Gaia Collaboration} et~al.,}{{Gaia
  Collaboration} et~al.}{2022}]{GaiaDR3_2022}
{Gaia Collaboration} et~al., 2022, \mn@doi [arXiv e-prints]
  {10.48550/arXiv.2208.00211}, \href
  {https://ui.adsabs.harvard.edu/abs/2022arXiv220800211G} {p. arXiv:2208.00211}

\bibitem[\protect\citeauthoryear{{Gieles} \& {Portegies Zwart}}{{Gieles} \&
  {Portegies Zwart}}{2011}]{Gieles_2011}
{Gieles} M.,  {Portegies Zwart} S.~F.,  2011, \mn@doi [\mnras]
  {10.1111/j.1745-3933.2010.00967.x}, \href
  {https://ui.adsabs.harvard.edu/abs/2011MNRAS.410L...6G} {410, L6}

\bibitem[\protect\citeauthoryear{{Gran} et~al.,}{{Gran}
  et~al.}{2022}]{Gran_2022}
{Gran} F.,  et~al., 2022, \mn@doi [\mnras] {10.1093/mnras/stab2463}, \href
  {https://ui.adsabs.harvard.edu/abs/2022MNRAS.509.4962G} {509, 4962}

\bibitem[\protect\citeauthoryear{{Hao}, {Xu}, {Wu}, {He}  \& {Bian}}{{Hao}
  et~al.}{2020}]{Hao_2020}
{Hao} C.,  {Xu} Y.,  {Wu} Z.,  {He} Z.,   {Bian} S.,  2020, \mn@doi [\pasp]
  {10.1088/1538-3873/ab694d}, \href
  {https://ui.adsabs.harvard.edu/abs/2020PASP..132c4502H} {132, 034502}

\bibitem[\protect\citeauthoryear{{Hao} et~al.,}{{Hao} et~al.}{2021}]{Hao_2021}
{Hao} C.~J.,  et~al., 2021, \mn@doi [\aap] {10.1051/0004-6361/202140608}, \href
  {https://ui.adsabs.harvard.edu/abs/Hao_2021} {652, A102}

\bibitem[\protect\citeauthoryear{{Hao}, {Xu}, {Wu}, {Lin}, {Liu}  \&
  {Li}}{{Hao} et~al.}{2022}]{Hao_2022}
{Hao} C.~J.,  {Xu} Y.,  {Wu} Z.~Y.,  {Lin} Z.~H.,  {Liu} D.~J.,   {Li} Y.~J.,
  2022, \mn@doi [\aap] {10.1051/0004-6361/202243091}, \href
  {https://ui.adsabs.harvard.edu/abs/2022A&amp;A...660A...4H} {660, A4}

\bibitem[\protect\citeauthoryear{{He}, {Xu}, {Hao}, {Wu}  \& {Li}}{{He}
  et~al.}{2021}]{He_2021}
{He} Z.-H.,  {Xu} Y.,  {Hao} C.-J.,  {Wu} Z.-Y.,   {Li} J.-J.,  2021, \mn@doi
  [Research in Astronomy and Astrophysics] {10.1088/1674-4527/21/4/93}, \href
  {https://ui.adsabs.harvard.edu/abs/2021RAA....21...93H} {21, 093}

\bibitem[\protect\citeauthoryear{{He} et~al.,}{{He} et~al.}{2022a}]{He_2022}
{He} Z.,  et~al., 2022a, \mn@doi [\apjs] {10.3847/1538-4365/ac5cbb}, \href
  {https://ui.adsabs.harvard.edu/abs/2022ApJS..260....8H} {260, 8}

\bibitem[\protect\citeauthoryear{{He}, {Wang}, {Luo}, {Li}, {Liu}  \&
  {Jiang}}{{He} et~al.}{2022b}]{He_2022_1}
{He} Z.,  {Wang} K.,  {Luo} Y.,  {Li} J.,  {Liu} X.,   {Jiang} Q.,  2022b,
  \mn@doi [\apjs] {10.3847/1538-4365/ac7c17}, \href
  {https://ui.adsabs.harvard.edu/abs/2022ApJS..262....7H} {262, 7}

\bibitem[\protect\citeauthoryear{{He}, {Luo}, {Wang}, {Ren}, {Peng}, {Cui},
  {Liu}  \& {Jiang}}{{He} et~al.}{2023a}]{He_2023_2}
{He} Z.,  {Luo} Y.,  {Wang} K.,  {Ren} A.,  {Peng} L.,  {Cui} Q.,  {Liu} X.,
  {Jiang} Q.,  2023a, \mn@doi [arXiv e-prints] {10.48550/arXiv.2305.10269},
  \href {https://ui.adsabs.harvard.edu/abs/2023arXiv230510269H} {p.
  arXiv:2305.10269}

\bibitem[\protect\citeauthoryear{{He}, {Liu}, {Luo}, {Wang}  \& {Jiang}}{{He}
  et~al.}{2023b}]{He_2023}
{He} Z.,  {Liu} X.,  {Luo} Y.,  {Wang} K.,   {Jiang} Q.,  2023b, \mn@doi
  [\apjs] {10.3847/1538-4365/ac9af8}, \href
  {https://ui.adsabs.harvard.edu/abs/2023ApJS..264....8H} {264, 8}

\bibitem[\protect\citeauthoryear{{Herschel}}{{Herschel}}{1786}]{Herschel1786}
{Herschel} W.,  1786, Philosophical Transactions of the Royal Society of London
  Series I, \href {https://ui.adsabs.harvard.edu/abs/1786RSPT...76..457H} {76,
  457}

\bibitem[\protect\citeauthoryear{{H{\o}g} et~al.,}{{H{\o}g}
  et~al.}{1997}]{Hog_1997}
{H{\o}g} E.,  et~al., 1997, \aap, \href
  {https://ui.adsabs.harvard.edu/abs/1997A&A...323L..57H} {323, L57}

\bibitem[\protect\citeauthoryear{{Huchra} \& {Geller}}{{Huchra} \&
  {Geller}}{1982}]{Huchra_1982}
{Huchra} J.~P.,  {Geller} M.~J.,  1982, \mn@doi [\apj] {10.1086/160000}, \href
  {https://ui.adsabs.harvard.edu/abs/1982ApJ...257..423H} {257, 423}

\bibitem[\protect\citeauthoryear{{Hunt} \& {Reffert}}{{Hunt} \&
  {Reffert}}{2021}]{Hunt_2021}
{Hunt} E.~L.,  {Reffert} S.,  2021, \mn@doi [\aap]
  {10.1051/0004-6361/202039341}, \href
  {https://ui.adsabs.harvard.edu/abs/2021A&amp;A...646A.104H} {646, A104}

\bibitem[\protect\citeauthoryear{{Hunt} \& {Reffert}}{{Hunt} \&
  {Reffert}}{2023}]{Hunt_2023}
{Hunt} E.~L.,  {Reffert} S.,  2023, \mn@doi [\aap]
  {10.1051/0004-6361/202346285}, \href
  {https://ui.adsabs.harvard.edu/abs/2023A&A...673A.114H} {673, A114}

\bibitem[\protect\citeauthoryear{Hunter et~al.}{Hunter
  et~al.}{2007}]{hunter_2007}
Hunter J.~D.,  et~al., 2007, Computing in science and engineering, 9, 90

\bibitem[\protect\citeauthoryear{{Jaehnig}, {Bird}  \&
  {Holley-Bockelmann}}{{Jaehnig} et~al.}{2021}]{Jaehnig_2021}
{Jaehnig} K.,  {Bird} J.,   {Holley-Bockelmann} K.,  2021, \mn@doi [\apj]
  {10.3847/1538-4357/ac1d51}, \href
  {https://ui.adsabs.harvard.edu/abs/2021ApJ...923..129J} {923, 129}

\bibitem[\protect\citeauthoryear{Jones, Oliphant, Peterson  et~al.}{Jones
  et~al.}{2001}]{Jones_2001}
Jones E.,  Oliphant T.,  Peterson P.,   et~al., 2001, {SciPy}: Open source
  scientific tools for {Python}, \url {http://www.scipy.org/}

\bibitem[\protect\citeauthoryear{{Kharchenko}, {Piskunov}, {Schilbach},
  {R{\"o}ser}  \& {Scholz}}{{Kharchenko} et~al.}{2012}]{Kharchenko_2012}
{Kharchenko} N.~V.,  {Piskunov} A.~E.,  {Schilbach} E.,  {R{\"o}ser} S.,
  {Scholz} R.~D.,  2012, \mn@doi [\aap] {10.1051/0004-6361/201118708}, \href
  {https://ui.adsabs.harvard.edu/abs/2012A&A...543A.156K} {543, A156}

\bibitem[\protect\citeauthoryear{{Kos} et~al.,}{{Kos} et~al.}{2018}]{Kos_2018}
{Kos} J.,  et~al., 2018, \mn@doi [\mnras] {10.1093/mnras/sty2171}, \href
  {https://ui.adsabs.harvard.edu/abs/2018MNRAS.480.5242K} {480, 5242}

\bibitem[\protect\citeauthoryear{{Kounkel}, {Covey}  \& {Stassun}}{{Kounkel}
  et~al.}{2020}]{Kounkel_2020}
{Kounkel} M.,  {Covey} K.,   {Stassun} K.~G.,  2020, \mn@doi [\aj]
  {10.3847/1538-3881/abc0e6}, \href
  {https://ui.adsabs.harvard.edu/abs/2020AJ....160..279K} {160, 279}

\bibitem[\protect\citeauthoryear{{Krone-Martins} \& {Moitinho}}{{Krone-Martins}
  \& {Moitinho}}{2014}]{Krone_2014}
{Krone-Martins} A.,  {Moitinho} A.,  2014, \mn@doi [\aap]
  {10.1051/0004-6361/201321143}, \href
  {https://ui.adsabs.harvard.edu/abs/2014A&A...561A..57K} {561, A57}

\bibitem[\protect\citeauthoryear{{Li} \& {Mao}}{{Li} \& {Mao}}{2023}]{Li_2023}
{Li} Z.,  {Mao} C.,  2023, \mn@doi [\apjs] {10.3847/1538-4365/acaf7d}, \href
  {https://ui.adsabs.harvard.edu/abs/2023ApJS..265....3L} {265, 3}

\bibitem[\protect\citeauthoryear{{Li} et~al.,}{{Li} et~al.}{2022}]{Li_2022}
{Li} Z.,  et~al., 2022, \mn@doi [\apjs] {10.3847/1538-4365/ac3c49}, \href
  {https://ui.adsabs.harvard.edu/abs/2022ApJS..259...19L} {259, 19}

\bibitem[\protect\citeauthoryear{{Liu} \& {Pang}}{{Liu} \&
  {Pang}}{2019}]{Liu_2019}
{Liu} L.,  {Pang} X.,  2019, \mn@doi [\apjs] {10.3847/1538-4365/ab530a}, \href
  {https://ui.adsabs.harvard.edu/abs/2019ApJS..245...32L} {245, 32}

\bibitem[\protect\citeauthoryear{Lloyd}{Lloyd}{1982}]{Lloyd_1982}
Lloyd S.,  1982, \mn@doi [IEEE Transactions on Information Theory]
  {10.1109/TIT.1982.1056489}, 28, 129

\bibitem[\protect\citeauthoryear{{Loktin} \& {Popova}}{{Loktin} \&
  {Popova}}{2017}]{Loktin_2017}
{Loktin} A.~V.,  {Popova} M.~E.,  2017, \mn@doi [Astrophysical Bulletin]
  {10.1134/S1990341317030154}, \href
  {https://ui.adsabs.harvard.edu/abs/2017AstBu..72..257L} {72, 257}

\bibitem[\protect\citeauthoryear{{Lynga}}{{Lynga}}{1987}]{Lynga1987}
{Lynga} G.,  1987, VizieR Online Data Catalog, \href
  {https://ui.adsabs.harvard.edu/abs/1995yCat.7092....0L} {p. VII/92A}

\bibitem[\protect\citeauthoryear{MacQueen}{MacQueen}{1967}]{MacQueen_1967}
MacQueen J.,  1967, Some methods for classification and analysis of
  multivariate observations, Proc. 5th {Berkeley} {Symp}. {Math}. {Stat}.
  {Probab}., {Univ}. {Calif}. 1965/66, 1, 281-297 (1967).

\bibitem[\protect\citeauthoryear{{Mermilliod}}{{Mermilliod}}{1995}]{Mermilliod1995}
{Mermilliod} J.-C.,  1995, in {Egret} D.,  {Albrecht} M.~A.,  eds, ~ Vol. 203,
  Information \& On-Line Data in Astronomy. p.~127,
  \mn@doi{10.1007/978-94-011-0397-8_12}

\bibitem[\protect\citeauthoryear{{Messier}}{{Messier}}{1774}]{Messier1771}
{Messier} C.,  1774, Mémoires de l'Académie Royale des Sciences, pp 435--461

\bibitem[\protect\citeauthoryear{{Ochsenbein}, {Bauer}  \&
  {Marcout}}{{Ochsenbein} et~al.}{2000}]{Ochsenbein_2000}
{Ochsenbein} F.,  {Bauer} P.,   {Marcout} J.,  2000, \mn@doi [\aaps]
  {10.1051/aas:2000169}, \href
  {http://cdsads.u-strasbg.fr/abs/2000A%26AS..143...23O} {143, 23}

\bibitem[\protect\citeauthoryear{{Parker}}{{Parker}}{2014}]{Parker_2014}
{Parker} R.~J.,  2014, in The Labyrinth of Star Formation. p.~431 (\mn@eprint
  {arXiv} {1208.0011}), \mn@doi{10.1007/978-3-319-03041-8_86}

\bibitem[\protect\citeauthoryear{Pedregosa et~al.,}{Pedregosa
  et~al.}{2011}]{scikit-learn}
Pedregosa F.,  et~al., 2011, Journal of Machine Learning Research, 12, 2825

\bibitem[\protect\citeauthoryear{{Pera}, {Perren}, {Moitinho}, {Navone}  \&
  {Vazquez}}{{Pera} et~al.}{2021}]{Pera_2021}
{Pera} M.~S.,  {Perren} G.~I.,  {Moitinho} A.,  {Navone} H.~D.,   {Vazquez}
  R.~A.,  2021, \mn@doi [\aap] {10.1051/0004-6361/202040252}, \href
  {https://ui.adsabs.harvard.edu/abs/2021A&A...650A.109P} {650, A109}

\bibitem[\protect\citeauthoryear{{Perren}, {V{\'a}zquez}  \& {Piatti}}{{Perren}
  et~al.}{2015}]{Perren_2015}
{Perren} G.~I.,  {V{\'a}zquez} R.~A.,   {Piatti} A.~E.,  2015, \mn@doi [\aap]
  {10.1051/0004-6361/201424946}, \href
  {https://ui.adsabs.harvard.edu/abs/2015A&A...576A...6P} {576, A6}

\bibitem[\protect\citeauthoryear{{Perryman} et~al.,}{{Perryman}
  et~al.}{1997}]{Perryman_1997}
{Perryman} M.~A.~C.,  et~al., 1997, \aap, \href
  {https://ui.adsabs.harvard.edu/abs/1997A&A...323L..49P} {323, L49}

\bibitem[\protect\citeauthoryear{Qin, Li, Chen  \& Zhong}{Qin
  et~al.}{2021}]{Qin_2021}
Qin S.-M.,  Li J.,  Chen L.,   Zhong J.,  2021, \mn@doi [Research in Astronomy
  and Astrophysics] {10.1088/1674-4527/21/2/45}, 21, 045

\bibitem[\protect\citeauthoryear{{Qin}, {Zhong}, {Tang}  \& {Chen}}{{Qin}
  et~al.}{2023}]{Qin_2023}
{Qin} S.,  {Zhong} J.,  {Tang} T.,   {Chen} L.,  2023, \mn@doi [\apjs]
  {10.3847/1538-4365/acadd6}, \href
  {https://ui.adsabs.harvard.edu/abs/2023ApJS..265...12Q} {265, 12}

\bibitem[\protect\citeauthoryear{Ripley}{Ripley}{1976}]{ripley_1976}
Ripley B.~D.,  1976, \mn@doi [Journal of Applied Probability]
  {10.2307/3212829}, 13, 255–266

\bibitem[\protect\citeauthoryear{Ripley}{Ripley}{1979}]{ripley_1979}
Ripley B.~D.,  1979, Journal of the Royal Statistical Society. Series B
  (Methodological), 41, 368

\bibitem[\protect\citeauthoryear{{Ryu} \& {Lee}}{{Ryu} \&
  {Lee}}{2018}]{Ryu_2018}
{Ryu} J.,  {Lee} M.~G.,  2018, \mn@doi [\apj] {10.3847/1538-4357/aab1ff}, \href
  {https://ui.adsabs.harvard.edu/abs/2018ApJ...856..152R} {856, 152}

\bibitem[\protect\citeauthoryear{{Santos-Silva} et~al.,}{{Santos-Silva}
  et~al.}{2021}]{Santos-Silva_2021}
{Santos-Silva} T.,  et~al., 2021, \mn@doi [\mnras] {10.1093/mnras/stab2409},
  \href {https://ui.adsabs.harvard.edu/abs/2021MNRAS.508.1033S} {508, 1033}

\bibitem[\protect\citeauthoryear{{Sim}, {Lee}, {Ann}  \& {Kim}}{{Sim}
  et~al.}{2019}]{Sim_2019}
{Sim} G.,  {Lee} S.~H.,  {Ann} H.~B.,   {Kim} S.,  2019, \mn@doi [Journal of
  Korean Astronomical Society] {10.5303/JKAS.2019.52.5.145}, \href
  {https://ui.adsabs.harvard.edu/abs/2019JKAS...52..145S} {52, 145}

\bibitem[\protect\citeauthoryear{{Skrutskie} et~al.,}{{Skrutskie}
  et~al.}{2006}]{Skrutskie_2006}
{Skrutskie} M.~F.,  et~al., 2006, \mn@doi [\aj] {10.1086/498708}, \href
  {https://ui.adsabs.harvard.edu/abs/2006AJ....131.1163S} {131, 1163}

\bibitem[\protect\citeauthoryear{{Tarricq}, {Soubiran}, {Casamiquela},
  {Castro-Ginard}, {Olivares}, {Miret-Roig}  \& {Galli}}{{Tarricq}
  et~al.}{2022}]{Tarricq_2022}
{Tarricq} Y.,  {Soubiran} C.,  {Casamiquela} L.,  {Castro-Ginard} A.,
  {Olivares} J.,  {Miret-Roig} N.,   {Galli} P.~A.~B.,  2022, \mn@doi [\aap]
  {10.1051/0004-6361/202142186}, \href
  {https://ui.adsabs.harvard.edu/abs/2022A&amp;A...659A..59T} {659, A59}

\bibitem[\protect\citeauthoryear{{Tian}}{{Tian}}{2020}]{Tian_2020}
{Tian} H.-J.,  2020, \mn@doi [\apj] {10.3847/1538-4357/abbf4b}, \href
  {https://ui.adsabs.harvard.edu/abs/Tian_2020} {904, 196}

\bibitem[\protect\citeauthoryear{{Tremmel} et~al.,}{{Tremmel}
  et~al.}{2013}]{Tremmel_2013}
{Tremmel} M.,  et~al., 2013, \mn@doi [\apj] {10.1088/0004-637X/766/1/19}, \href
  {https://ui.adsabs.harvard.edu/abs/2013ApJ...766...19T} {766, 19}

\bibitem[\protect\citeauthoryear{Van Der~Walt, Colbert  \& Varoquaux}{Van
  Der~Walt et~al.}{2011}]{vanDerWalt_2011}
Van Der~Walt S.,  Colbert S.~C.,   Varoquaux G.,  2011, Computing in Science \&
  Engineering, 13, 22

\bibitem[\protect\citeauthoryear{{Vasiliev} \& {Baumgardt}}{{Vasiliev} \&
  {Baumgardt}}{2021}]{Vasiliev_2021}
{Vasiliev} E.,  {Baumgardt} H.,  2021, \mn@doi [\mnras]
  {10.1093/mnras/stab1475}, \href
  {https://ui.adsabs.harvard.edu/abs/2021MNRAS.505.5978V} {505, 5978}

\bibitem[\protect\citeauthoryear{{Zari}, {Hashemi}, {Brown}, {Jardine}  \& {de
  Zeeuw}}{{Zari} et~al.}{2018}]{Zari_2018}
{Zari} E.,  {Hashemi} H.,  {Brown} A.~G.~A.,  {Jardine} K.,   {de Zeeuw} P.~T.,
   2018, \mn@doi [\aap] {10.1051/0004-6361/201834150}, \href
  {https://ui.adsabs.harvard.edu/abs/Zari_2018} {620, A172}

\bibitem[\protect\citeauthoryear{{van Rossum}}{{van
  Rossum}}{1995}]{vanRossum_1995}
{van Rossum} G.,  1995, Report CS-R9526, {Python} tutorial.
pub-CWI, pub-CWI:adr

\makeatother
\end{thebibliography}



\appendix

\section{Databases cleaning}
\label{app:db_cleaning}

We describe here a summary of the cleaning and standardizing processes applied
on almost all of the catalogues mentioned in Table~\ref{tab:references}. Most of
these are small tasks, like manually editing names so that they are
consistent across databases, but they were necessary (and rather time
consuming). For convenience we abbreviate \cite{Kharchenko_2012} as KHAR12 and
\cite{Bica_2019} as BICA19.
The catalogue used for identifying globular clusters (GCs) is that
of~\cite{Vasiliev_2021} with the addition of the Gran 2, 3, 4, and 5 objects
from~\cite{Gran_2022}.\\

\noindent\cite{Kharchenko_2012}: Selected all entries with \emph{class} equal to
``OPEN STAR CLUSTER'', resulting in 2858 entries. The astrometry values in this
database are of low quality, we do not use these values in the
membership estimation process.
Added the preferred denomination VDBH to vdBergh-Hagen and VDB to vdBergh
clusters, per CDS recommendation (this is done across all the catalogues).
Removed entries that match  Galactic clusters: ESO 456-29 (Gran 1), FSR 1716,
FSR 1758, VDBH 140.\\

\noindent\cite{Loktin_2017}: Many proper motion values in this catalogue are clearly
wrong (e.g: the values for NGC 2516). We thus do not use these values in the
membership estimation process.
Removed entries that match  Galactic clusters: Berkeley 42 (NGC 6749), Lynga 7
(BH 184).
Fixed the name of four listed OCs: Sauer5 $\rightarrow$ Saurer 5,
Teusch61 $\rightarrow$ Teutsch 61, AlessiJ2327+55$\rightarrow$ Alessi
J2327.0+55, Sigma Ori $\rightarrow$ Sigma Orionis.\\

\noindent\cite{Bica_2019}: The Vizier table contain 10978 entries, we keep only those
with class OC (open cluster) or OCC (open cluster candidate). This
reduces the list to 3564 entries.

This is the only DB that lists the~\cite{Ryu_2018} clusters. The original
article claims to have found 721 new OCs (923 minus 202 embedded). BICA19 (page
11) says that the Ryu \& Lee article lists 719 OCs (921 minus 202 embedded).
BICA19 lists in its Vizier table only 711 Ryu OCs, 4 of which are listed with
alternative names (Teutsch J1814.6-2814 $\rightarrow$ Ryu 563,
Quartet $\rightarrow$ Ryu 858, ``GLIMPSE 70, Mercer 70'' $\rightarrow$ Ryu 273,
``LS 468|La Serena 468'' $\rightarrow$ Ryu 094). Hence there are
707 Ryu clusters in the final BICA19 Vizier table.
Removed entries: MWSC 2776, FSR 523, FSR 847, FSR 436 (listed twice, removed
one of the entries); ESO 393-3 (listed twice and no available data in CDS to
decide, removed both); MWSC 1025, 1482, 948, 3123, 1997, 1840, 442, 1808, 2204
(listed twice, removed name from both entries as they were not found in KHAR12);
ESO 97-2 (removed from Loden 848 as it matches the position of Loden 894);
FSR 972, OCL 344, Collinder 384, FSR 179 (listed twice, removed from both
entries as we found no available data in CDS to check);
MWSC 206 (listed twice, removed the entry that also showed FSR 60 since the
coordinates for FSR 60 are a better match in KHAR12 for the entry with the
single FSR 60 name); Alessi J0715.6-0722 (removed as its position matches that
of Alessi J0715.6-0727).

Fixed the name of the following entries: FSR 429.MWSC 3667 $\rightarrow$ FSR
429,MWSC 3667; Carraro 1.MWSC 1829 $\rightarrow$ Carraro 1,MWSC 1829;
Cernik 39 $\rightarrow$ Czernik 39; de Wit 1 $\rightarrow$ Wit 1 (to match KHAR12);
JS 1 $\rightarrow$ Juchert-Saloran 1 (to match KHAR12);
ESO 589-26,MW $\rightarrow$ ESO 589-26; Alessi J2327.6+5535 $\rightarrow$ Alessi
J2327.0+55; TRSG 1 $\rightarrow$ RSG 1; Dol-Dzim 9 added DoDz 9 (to match
KHAR12); Dol-Dzim 11 added DoDz 11 (to match KHAR12).
Removed entries that match GCs: ESO 456-29,MWSC 2761 (Gran 1);
ESO 93-8,MWSC 1932; FSR 1758,MWSC 2617; VDBH 140,vdBergh-Hagen 140,FSR 1632,MWSC
2071.\\

\noindent\cite{Sim_2019}: Added a 'plx' column estimated as the inverse of the distance 
(the distance in parsecs is included in the catalogue).\\

\noindent\citet[][LIU19]{Liu_2019}: Added the identifier 'FoF' to all the entries to
match HUNT23. Changed 'LP' for 'FoF' in all the catalogues where it appeared,
for consistency.\\

\noindent\cite{Cantat-Gaudin_2020}: Changed Sigma Ori $\rightarrow$ Sigma Orionis.\\

\noindent\cite{Castro-Ginard_2020}: Fixed wrong right ascension value for UBC
595 and UBC 181.\\

\noindent\cite{Hao_2020}: Added the acronym 'HXWHB' to match HUNT23.\\

\noindent\cite{He_2021}: Added 'CWNU' acronym for consistency across catalogues.\\

\noindent\cite{Dias_2021}: Lists 177 LIU19 clusters because it includes
clusters not listed as new by the authors. We remove all except those listed as
new in LIU19. Cluster LP 866 was duplicated (listed also as LP 0866), entry was
removed. Changed Sigma Ori $\rightarrow$ Sigma Orionis.\\

\noindent\cite{Hao_2021}: This database contains dozens of duplicated entries and even
some that are listed thrice, e.g: ESO 130-06, ESO 368-11, ESO 368-14 and
Basel 11a. Furthermore, duplicated clusters are assigned very
different fundamental parameters (e.g., Alessi 44 is listed twice and assigned
logarithmic ages of 7.82 and 8.42). Of the almost 4000 listed OCs, $\sim$15\%
show a difference in the mean parallax value with those from CANTAT20 larger
than 50\%. Finally some clusters have wildly incorrect astrometric parameters.
For example the cluster Melotte 25 is assigned a parallax of 0.264 mas in this
catalogue when its true value is larger than 21 mas. We thus unfortunately
decided to exclude this catalogue from our list.\\

\noindent\cite{Hao_2022}: Removed the listed entry OC 0586 as a duplicate of the GC BH
140.\\

\noindent\cite{Hunt_2023}: Removed GCs listed as OCs: Palomar 2, 7 (listed
as IC 1276) 8, 10, 11, 12; ESO 452-11 (1636-283); Pismis 26 (Ton 2); Lynga 7 (BH
184). New candidates HSC 2890 and HSC 134 were removed as their position and
astrometry match those of the GCs Gran 3 and 4. Candidate HSC 2605
has very similar coordinates and proper motions to GC NGC 5139 but its
parallax is different, so it was not removed.
Removed moving groups and Theia objects from~\cite{Kounkel_2020}.
Fixed: ESO 429-429 $\rightarrow$ ESO 429-02 (position corresponds to this OC),
AH03 J0748+26.9 $\rightarrow$ AH03 J0748-26.9,
Juchert J0644.8+0925 $\rightarrow$ Juchert J0644.8-0925,
Teutsch J0718.0+1642 $\rightarrow$ Teutsch J0718.0-1642,
Teutsch J0924.3+5313 $\rightarrow$ Teutsch J0924.3-5313,
Teutsch J1037.3+6034 $\rightarrow$ Teutsch J1037.3-6034,
Teutsch J1209.3+6120 $\rightarrow$ Teutsch J1209.3-6120,
Collinder 302 changed position to (RA=246.525, DEC=-26.233), it was centred
on GC NGC 6121.

For $\sim$160 HSC candidates we updated their centre values in coordinates.
These are extended and irregular objects for which the median positions of their
members was more than 1 deg away from the stored values in the HUNT23
database.\\

\noindent\cite{Li_2023}: Database lists 56 'LISC' clusters but only 35 are kept as real
objects. The parallax distances are in very bad agreement with the the estimated
distance moduli. HUNT23 recovers 0\% of these clusters. We keep the catalogue
but advise caution.\\

\noindent\cite{Chi_2023}: The article mentions 83 clusters but only 82 are visible in the
article table that lists them. No Vizier data was available at the moment of
writing this article and no answer was received after enquiring the author.
answer. Added 'LISC-III' to the names to match HUNT23.


\bsp	
\label{lastpage}
\end{document}